\documentclass[letterpaper,titlepage,11pt]{article}

\pdfoutput=1

\usepackage{amsmath,amssymb,amsthm,mathrsfs,bbm}
\usepackage{latexsym,amscd,amsbsy,amsfonts,dsfont}
\usepackage{graphicx}
\usepackage{xcolor}
\usepackage[
      colorlinks=true,
      linkcolor=blue,
      urlcolor=blue,
      filecolor=black,
      citecolor=red,
      pdfstartview=FitV,
      pdftitle={},
        pdfauthor={Roberto Emparan, Antonia M Frassino, Benson Way},
        pdfsubject={},
        pdfkeywords={},
        pdfpagemode=None,
        bookmarksopen=true
      ]{hyperref}\usepackage{caption}

\usepackage[utf8]{inputenc}
\usepackage{multirow}
\usepackage{makecell}

\newcommand{\beq}{\begin{equation}}
\newcommand{\eeq}{\end{equation}}
\newcommand{\beqa}{\begin{eqnarray}}
\newcommand{\eeqa}{\end{eqnarray}}
\newcommand{\bea}{\begin{eqnarray}}
\newcommand{\eea}{\end{eqnarray}}

\newcommand{\nn}{\nonumber}

\newcommand{\lp}{\left(}
\newcommand{\rp}{\right)}

\newcommand{\ord}[1]{{\mathcal O}\lp #1\rp}

\newcommand{\Z}{\mathbb{Z}}

\newcommand{\R}{\mathbb{R}}

\def\clock{{\count0=\time
           \divide\count0 60
           \ifnum\count0<10 0\fi\the\count0
           \multiply\count0 -60 \advance\count0 \time
           :\ifnum\count0<10 0\fi \the\count0
         }}
\newcommand{\timestamp}{{\small\vbox{\hbox{\tt\jobname.tex}
\hbox{\the\day/\the\month/\the\year, \clock}}}}

\numberwithin{equation}{section}

\begin{document}

\begin{titlepage}
\leftline{}
\vskip 2cm
\centerline{\LARGE \bf Quantum BTZ black hole}
\bigskip

\vskip 1.2cm
\centerline{\bf Roberto Emparan$^{a,b}$, Antonia Micol Frassino$^{b}$, Benson Way$^{b}$
}

\vskip 0.5cm
\centerline{\sl $^{a}$Instituci\'o Catalana de Recerca i Estudis
Avan\c cats (ICREA)}
\centerline{\sl Passeig Llu\'{\i}s Companys 23, E-08010 Barcelona, Spain}
\smallskip
\centerline{\sl $^{b}$Departament de F{\'\i}sica Qu\`antica i Astrof\'{\i}sica, Institut de
Ci\`encies del Cosmos,}
\centerline{\sl  Universitat de
Barcelona, Mart\'{\i} i Franqu\`es 1, E-08028 Barcelona, Spain}
\smallskip

\vskip 0.5cm
\centerline{\small\tt emparan@ub.edu, antoniam.frassino@icc.ub.edu, benson@icc.ub.edu}

\vskip 1.cm
\centerline{\bf Abstract} \vskip 0.2cm \noindent

We study a holographic construction of quantum rotating BTZ black holes that incorporates the exact backreaction from strongly coupled quantum conformal fields. It is based on an exact four-dimensional solution for a black hole localized on a brane in AdS$_4$, first discussed some years ago but never fully investigated in this manner. Besides quantum CFT effects and their backreaction, we also investigate the role of higher-curvature corrections in the effective three-dimensional theory. We obtain the quantum-corrected geometry and the renormalized stress tensor. We show that the quantum black hole entropy, which includes the entanglement of the fields outside the horizon, satisfies the first law of thermodynamics exactly, even in the presence of backreaction and with higher-curvature corrections, while the Bekenstein-Hawking-Wald entropy does not. This result, which involves a rather non-trivial bulk calculation, shows the consistency of the holographic interpretation of braneworlds. We compare our renormalized stress tensor to results derived for free conformal fields, and for a previous holographic construction without backreaction effects, which is shown to be a limit of the solutions in this article.

\noindent

\end{titlepage}
\pagestyle{empty}
\small

\addtocontents{toc}{\protect\setcounter{tocdepth}{2}}

\tableofcontents
\normalsize
\newpage
\pagestyle{plain}
\setcounter{page}{1}

\section{Introduction}

Despite the lack of a precise definition of a quantum black hole within a complete quantum theory of gravity, one can still gain insight through semi-classical approximations.  One sensible approach is to treat gravity classically while fully accounting for the backreaction of all other quantum fields. This is the study of the `semi-classical Einstein equations'
\beq\label{semieins}
G_{\mu\nu}(g_{\alpha\beta})=8\pi G \langle T_{\mu\nu}(g_{\alpha\beta}) \rangle\,,
\eeq
where $G_{\mu\nu}$ is the gravitational Einstein tensor (possibly with a cosmological constant) for a spacetime metric $g_{\alpha\beta}$, and $\langle T_{\mu\nu} \rangle$ is the renormalized stress tensor of the (non-gravitational) quantum matter fields in that spacetime. Quantum fluctuations of the metric can be comparatively suppressed by including a large number of matter degrees of freedom.

Ideally, one would simultaneously solve for both the coupled system of the metric $g_{\alpha\beta}$ and the correlation functions of quantum field operators.  However, this problem is often intractable.  Instead, the backreaction effects are typically assumed to be small and the problem is approached perturbatively.  Nevertheless, there are non-perturbative approaches available in special cases.  The complete backreaction problem \eqref{semieins} has been solved in some two-dimensional models \cite{Fabbri:2005mw,Callan:1992rs,Almheiri:2019psf}, while in more dimensions it can be tackled through a holographic reformulation.  In this article, we shall use the holographic approach to exactly solve a variant of \eqref{semieins} to find the quantum form of the three-dimensional BTZ black hole \cite{Banados:1992wn,Banados:1992gq}---quBTZ, for short.

The AdS$_4$/CFT$_3$ duality maps the quantum theory of three-dimensional conformal fields to a problem of gravitational dynamics in a four-dimensional AdS bulk spacetime.  In the large $N$ expansion of the CFT, the leading order (planar-diagram limit) is dual to classical gravitational bulk physics. One variation of this duality enables the study of the CFT in a dynamical spacetime by introducing a brane in the bulk.  This setup is similar to a Randall-Sundrum construction \cite{Randall:1999vf}, but is more precisely described as a Karch-Randall model with AdS$_3$ branes \cite{Karch:2000ct}. The problem \eqref{semieins} is now in the form of an effective gravitational theory
\beq\label{holobackr}
\mathcal{G}_{\mu\nu}(g_{\alpha\beta})=8\pi G \langle T_{\mu\nu}(g_{\alpha\beta}) \rangle_\textrm{planar}\,,
\eeq
where $\mathcal{G}_{\mu\nu}$ includes higher-curvature corrections, which can be reduced via holography to
solving the classical gravitational equations of a braneworld model in one more dimension \cite{Verlinde:1999fy,Gubser:1999vj}. Good recent discussions of this duality and its subtleties can be found in \cite{Geng:2020qvw,Chen:2020uac}. One may include non-planar CFT corrections to \eqref{holobackr} by computing bulk quantum effects, in a perturbative expansion that resembles but is not the same as the more conventional perturbative backreaction approach to \eqref{semieins}.\footnote{It is in fact the conventional perturbative backreaction problem in the four-dimensional bulk, but not in the three-dimensional boundary dual.}

Ref.~\cite{Emparan:2002px} used these ideas, and the exact construction in \cite{Emparan:1999fd} of black holes localized on a brane in AdS$_4$, to present a holographic solution to \eqref{holobackr} for the static, non-rotating quantum BTZ black hole. However, the analysis of the solutions in \cite{Emparan:2002px} was incomplete,\footnote{In \cite{Emparan:2002px} the quBTZ black hole was one among other solutions used for a different purpose than our main motivation here. See also \cite{Tanaka:2002rb}.}  and neither \cite{Emparan:2002px} nor \cite{Emparan:1999fd} discussed how backreaction effects are extracted nor took proper account of the higher-curvature corrections in the effective three-dimensional theory. Motivated by this and by later developments, we are led to revisit and reassess the quantum properties of the BTZ black hole.

\medskip

\noindent\textit{The renormalized quantum stress tensor.} One of our aims is to extend, relate, and compare different calculations of the renormalized stress tensor of conformal fields in the rotating BTZ black hole and its backreaction. The majority of previous works \cite{Steif:1993zv,Shiraishi:1993nu,Shiraishi:2018pdw,Lifschytz:1993eb,Martinez:1996uv,Casals:2016odj,Casals:2019jfo} study free conformal fields, but \cite{Hubeny:2009rc} used holography to obtain the stress tensor of a strongly coupled CFT in the rotating BTZ geometry (without backreaction), using a bulk construction apparently very different than the one in \cite{Emparan:2002px,Emparan:1999fd}. In this article we extend the static construction in \cite{Emparan:2002px} to the richer general solution of \eqref{holobackr} for the rotating quantum BTZ black hole. Then we compare the results to previous calculations of the quantum stress tensor, both for holographic and free CFTs. We show that the holographic bulk solution in \cite{Hubeny:2009rc} arises as a limit of ours, and we present the correct result for the stress tensor in the presence of rotation.\footnote{We have communicated with the authors of \cite{Hubeny:2009rc}, who agree with our conclusions.} The comparison between the holographic and free CFT calculations reveals similarities that are not a direct consequence of conformal symmetry, but also significant qualitative differences. In general the holographic result is considerably simpler.

\medskip

\noindent\textit{Quantum CFT backreaction and higher-curvature corrections.} The holographic interpretation of the bulk geometry as a solution to \eqref{holobackr} incorporates in an exact manner two kinds of quantum modifications of the BTZ black hole. Recall that the effective three-dimensional theory comes with a cutoff. The quantum CFT degrees of freedom with energies above the cutoff give rise to the `induced gravity' on the brane, which includes the higher-curvature corrections to the gravitational theory in the left-hand side of \eqref{holobackr}, and these modify the BTZ solution. The low-energy degrees of freedom of the quantum CFT, instead, directly backreact on the geometry by entering in the right-hand side of \eqref{holobackr}.

Both the low- and high-energy quantum effects are incorporated in the bulk construction in an exact manner (in the planar limit), but for small backreaction their imprints on the geometry can be easily separated as being,  respectively, of linear and quadratic order in the strength of the backreaction.  The small backreaction parameter is $c\, G_3/L_3$, where $c$ is the central charge of the CFT and $G_3$ and $L_3$ are the effective three-dimensional Newton's constant and AdS$_3$ radius. The low-energy, direct backreaction effects are proportional to the number of quantum degrees of freedom, $c$, and hence are linear in $c\, G_3/L_3$. On the other hand, the curvature corrections appear at quadratic order in $c\, G_3/L_3$, since the cutoff length of the effective theory is proportional to $c\, G_3$ and the leading curvature corrections are quadratic in length. Therefore, at linear order, we can cleanly extract the leading CFT backreaction effects while ignoring higher-curvature modifications. At higher orders, the effects mix, but the corrections can be systematically studied in a manner that is much simpler than in the conventional approach. We find that the higher-order backreaction in our solutions has an intriguing simplicity. The most remarkable exact results appear in the study of quantum thermodynamics.

\medskip

\noindent\textit{Quantum entropy and the first law.} A main goal of our work, largely unexplored in this context, is to investigate the entropy of the quantum black hole and its thermodynamic properties. In a solution to \eqref{semieins}, we expect that this is a `generalized entropy' consisting of two terms, namely
\beq\label{quS}
S_\text{gen}=\frac{A}{4G}+S_\text{out}\,,
\eeq
where $A$ is the area of the horizon of the black hole and $S_\text{out}$ is the entanglement entropy of the quantum fields in the region outside the black hole, after absorbing the leading (divergent) contribution $\propto A$ in a renormalization of $G$. More generally, with higher-curvature gravitational terms in \eqref{holobackr}, the Bekenstein-Hawking entropy $A/4G$ is replaced by the Wald entropy \cite{Wald:1993nt}. By the arguments above, the quantum CFT entropy $S_\text{out}$ is proportional to $c$ and hence distinct from the leading curvature corrections in the Wald entropy, which are quadratic in the small parameter $c\, G_3/L_3$

The holographic approach directly computes $S_\text{gen}$ from the horizon area in the bulk geometry,
\beq\label{bulkS}
S_\text{gen}=\frac{A_\text{bulk}}{4G_\text{bulk}}\,,
\eeq
which in general is different than the Bekenstein-Hawking-Wald entropy for the brane black hole, since $A$ and $G$ (and the Wald correction terms) are quantities defined and measured on the brane. The difference between the two entropies is interpreted holographically as $S_\text{out}$.
This is in fact an application to braneworld holography of the Ryu-Takayanagi formula for the entanglement entropy of quantum fields \cite{Ryu:2006bv}, first considered in this context in \cite{Emparan:2006ni}. In this interpretation, the horizon in the bulk is an RT minimal surface and therefore all of $S_\text{gen}$, and not only $S_\text{out}$, must be regarded as entanglement entropy. The reason is that gravity on the brane---not only the higher-curvature terms, but also the Einstein-Hilbert term in the action---is induced by integrating the ultraviolet degrees of freedom of the CFT. Therefore, the Bekenstein-Hawking-Wald entropy must be seen as fully induced by the entanglement across the horizon of very short wavelength quantum fluctuations. Hence, in the braneworld the entanglement entropies of the quantum degrees of freedom above and below the cutoff show up as the two distinct terms in \eqref{quS}. These same ideas imply that, within this set up, the bulk RT surface can be used to find the quantum extremal surface \cite{Engelhardt:2014gca} for a system on the brane, an idea that recently has been used to good effect in \cite{Almheiri:2019hni,Almheiri:2019psy}.

We will show that the entropy of the quantum black hole, $S_\text{gen}$ \eqref{bulkS}, does behave like a thermodynamic entropy, in that it satisfies the first law of quantum black holes
\beq\label{q1stlaw}
TdS_\text{gen} =dM -\Omega d J
\eeq
where $M$, $J$, $T$ and $\Omega$ are magnitudes of the black hole that are all measured on the brane. In constrast, the Bekenstein-Hawking-Wald entropy does not satisfy the first law when the CFT backreaction is included. Proofs that it does \cite{Wald:1993nt} assume that the entropy of matter and radiation is negligible compared to the gravitational black hole entropy, but our construction includes both of them.

This is a non-trivial test of the holographic interpretation of braneworlds, since it is not obvious that \eqref{q1stlaw} must hold, given that the quantities on the left and right side of it belong in different worlds. The correct entropy for the first law is given by the bulk horizon area $A_\text{bulk}$, 
but the mass and spin are defined and measured as magnitudes on the brane.  As we will see, the presence of higher-curvature corrections and the global structure of the rotating bulk solution make the calculation of all the magnitudes delicate, so their apparent conspiracy to yield \eqref{q1stlaw} is remarkable.

Even more impressively, we find that \eqref{q1stlaw} holds exactly to all orders in the backreaction and higher-curvature corrections. In order to understand what this means, note that $S_\text{gen}$ is defined as an exact magnitude of the bulk solution (up to quantum bulk corrections), independently of the effective three-dimensional theory. On the other hand, $M$ and $J$ are three-dimensional magnitudes, and their definition in terms of the 3D metric coefficients receives corrections order by order in the curvature expansion of the effective theory. We will prove that \eqref{q1stlaw} holds not only when we account for quadratic curvature corrections; beyond this order, with a simple and natural all-order resummation for the exact values of $M$ and $J$, the first law \eqref{q1stlaw} is exactly satisfied.

One may surmise that the explanation of this exact result is that \eqref{q1stlaw} is nothing but the classical `first law in the bulk' for our four-dimensional solution. However, finding this bulk law is not without difficulties. On the one hand, the black hole is accelerating in the bulk and the bulk asymptotics is non-standard due to the presence of the brane. On the other hand, even after these ambiguities are fixed, it is not obvious that the bulk-defined $M$ and $J$ should exactly agree with the three-dimensional ones that enter in \eqref{q1stlaw}, furthermore including higher-curvature corrections. We will return to this point in the final section.\footnote{Ref.~\cite{Emparan:1999fd}, and also \cite{Emparan:1999wa}, did a calculation equivalent to proving \eqref{q1stlaw} for the static black hole. It was  interpreteted as defining $M$ as a `bulk mass', but the identification as the three-dimensional mass in \cite{Emparan:1999fd} was not consistent beyond the leading order of the effective theory.}

With this result, the holographic interpretation of \eqref{quS} and \eqref{bulkS}, and more generally the holographic dual interpretation \eqref{holobackr} for the braneworld theory, are shown to be consistent with basic thermodynamics.

\medskip

The outline of the paper is as follows. In the next section, we describe the bulk construction of the holographic quantum BTZ black hole in the static case. We discuss the effective three-dimensional theory, including the leading higher-curvature corrections. We obtain the CFT stress tensor, study its properties in comparison with other calculations, and investigate the quantum entropy and the first law. In section~\ref{sec:rotquBTZ} we extend the study to the rather more complex rotating quantum black hole. Sec.~\ref{sec:discuss} ends with a discussion of remaining issues and further open ideas. Appendix \ref{app:list} is a glossary of symbols. In appendix \ref{app:zerobr}, we prove that the holographic construction in \cite{Hubeny:2009rc} is recovered as a limit of the one in this paper, and in appendix \ref{app:TabBTZ} we give the holographic renormalized stress tensor in another form.

\section{Holographic dual of static quBTZ}

As explained in \cite{Emparan:2002px,Tanaka:2002rb}, the holographic method of solving \eqref{holobackr} for a quantum-corrected black hole is through a classical bulk dual with a black hole localized on a braneworld. Refs.~\cite{Emparan:1999wa,Emparan:1999fd} presented exact solutions for black holes on three-dimensional branes based on the AdS$_4$ C-metric. Our approach will be based on these same solutions, now holographically interpreted.

We will begin with the simpler case of static, non-rotating solutions. Although parts of the technical analysis were done in \cite{Emparan:1999fd,Emparan:2002px}, we will sometimes present a different interpretation of results and provide a more detailed investigation. Two important aspects that were not discussed earlier are the effects of the strength of backreaction, and the higher-curvature corrections in the effective theory.

\subsection{The AdS C-Metric}

The AdS$_4$ C-metric, which is a central element of our construction, is part of the Pleba{\'n}ski-Demia{\'n}ski family of type D metrics \cite{Plebanski:1976gy}, a remarkably versatile class of exact solutions that have found many different applications. The C-metric has been presented in a variety of coordinates and parametrizations, and we will choose one that is particularly well suited for studying the black hole on the brane, i.e., the holographic quantum black hole. We write the metric in the form
\beq\label{statc}
ds^2=\frac{\ell^2}{\lp \ell+xr\rp^2}\lp -H(r) dt^2+\frac{dr^2}{H(r)}
+r^2\lp \frac{dx^2}{G(x)}+G(x)d\phi^2\rp\rp\,,
\eeq
where
\beqa\label{HG}
H(r)&=& \frac{r^2}{\ell_3^2}+\kappa -\frac{\mu\ell}{r}\,,\\
G(x)&=& 1-\kappa x^2-\mu x^3\,.
\eeqa
Here, $\kappa$, $\mu$, $\ell$ and $\ell_3$ are parameters whose meaning we will soon clarify.\footnote{Compared to \cite{Emparan:1999fd}, \eqref{statc} is obtained by making $\lambda=(\ell/\ell_3)^2$, $A=1/\ell$,  $k=-\kappa$, $2mA=\mu$, $y=-\ell/r$, and rescaling $t\to t/\ell$.\label{transl}}

The metric \eqref{statc} is a solution to the Einstein equations
\beq
R_{ab}=-3\lp \frac1{\ell^2}+\frac1{\ell_3^2}\rp g_{ab}\,,
\eeq
so the AdS$_4$ radius is
\beq\label{l34}
\ell_4=\lp \frac1{\ell^2}+\frac1{\ell_3^2}\rp^{-1/2}\,.
\eeq
We will see presently that the length scale $\ell_3$ is the AdS$_3$ radius on the brane. It is possible to eliminate $\ell$ and use $\ell_3$ and $\ell_4$ as parameters, but as we shall see, $\ell$ is directly related to the brane tension and the strength of backreaction in the dual theory.  For this reason, we will mostly work with $\ell$ and $\ell_3$. While solutions with imaginary $\ell$ are in principle valid,\footnote{Imaginary $\ell$ would be appropriate for deSitter branes.}  we will take it to be real and (without loss of generality) non-negative,
\beq
0\leq \ell<\infty\,,
\eeq
so that $\ell_3 >\ell_4$.

The other parameters in the solution are dimensionless: a discrete one,
\beq
\kappa=\pm 1, 0\,,
\eeq
and the non-negative real number $\mu$. We will be mostly interested in the case $\kappa=-1$, since this is needed to have BTZ on the brane, but we will carry out the study with arbitrary $\kappa$ to include other interesting quantum black holes. Solutions with $\kappa=0$ need not be considered separately since they are recovered in the limit $\mu\to\infty$ of the other two cases. We will eventually see from later results that $\mu>0$ accounts for the holographic quantum corrections to the black hole.

In our investigation of the induced three-dimensional physics, we will typically be interested in keeping $\ell_3$ fixed and then study the solutions for different values of $\mu$ and $\ell$ (i.e the dimensionless quantity $\ell/\ell_3$). Then $\ell_4$ is a derived scale, as befits the notion that the bulk emerges from boundary physics.

\subsection{Karch-Randall-Sundrum braneworld holography}

Let us first get some intuition for the metric by setting $\mu=0$ and noting that if we change coordinates $(x,r)\to(\sigma,\hat{r})$ with
\beq\label{changeco}
\cosh\sigma=\frac{\ell_3}{\ell_4}\frac{\sqrt{1+\frac{r^2 x^2}{\ell_3^2}}}{\left|1+\frac{r x}{\ell}\right|}\,,\qquad
\hat{r}=r\sqrt{\frac{1-\kappa x^2}{1+\frac{r^2 x^2}{\ell_3^2}}}\,,
\eeq
then the geometry becomes more explicitly pure AdS$_4$,
\beq\label{emptyads4}
ds^2=\ell_4^2 d\sigma^2+\frac{\ell_4^2}{\ell_3^2}\cosh^2\sigma \lp \frac{d\hat{r}^2}{\frac{\hat{r}^2}{\ell_3^2}+\kappa}-\lp\frac{\hat{r}^2}{\ell_3^2}+\kappa\rp dt^2+\hat{r}^2 d\phi^2\rp\,,
\eeq
in a foliation by constant $\sigma$ slices that are AdS$_3$ with radius $\ell_4\cosh\sigma$. Each value of $\kappa$ gives slices with a recognizable form of AdS$_3$: global, Poincar\'e, or BTZ, for $\kappa=+1,0,-1$ respectively. If we cut the $\kappa=+1$ geometry with a brane at constant $\sigma=\sigma_b$ (with AdS$_3$ radius $\ell_4\cosh\sigma_b$) and discard the region $\sigma>\sigma_b$, the construction gives the ground state of the Karch-Randall set up.

\begin{figure}[t!h]
\centering
\includegraphics[width=.49\textwidth]{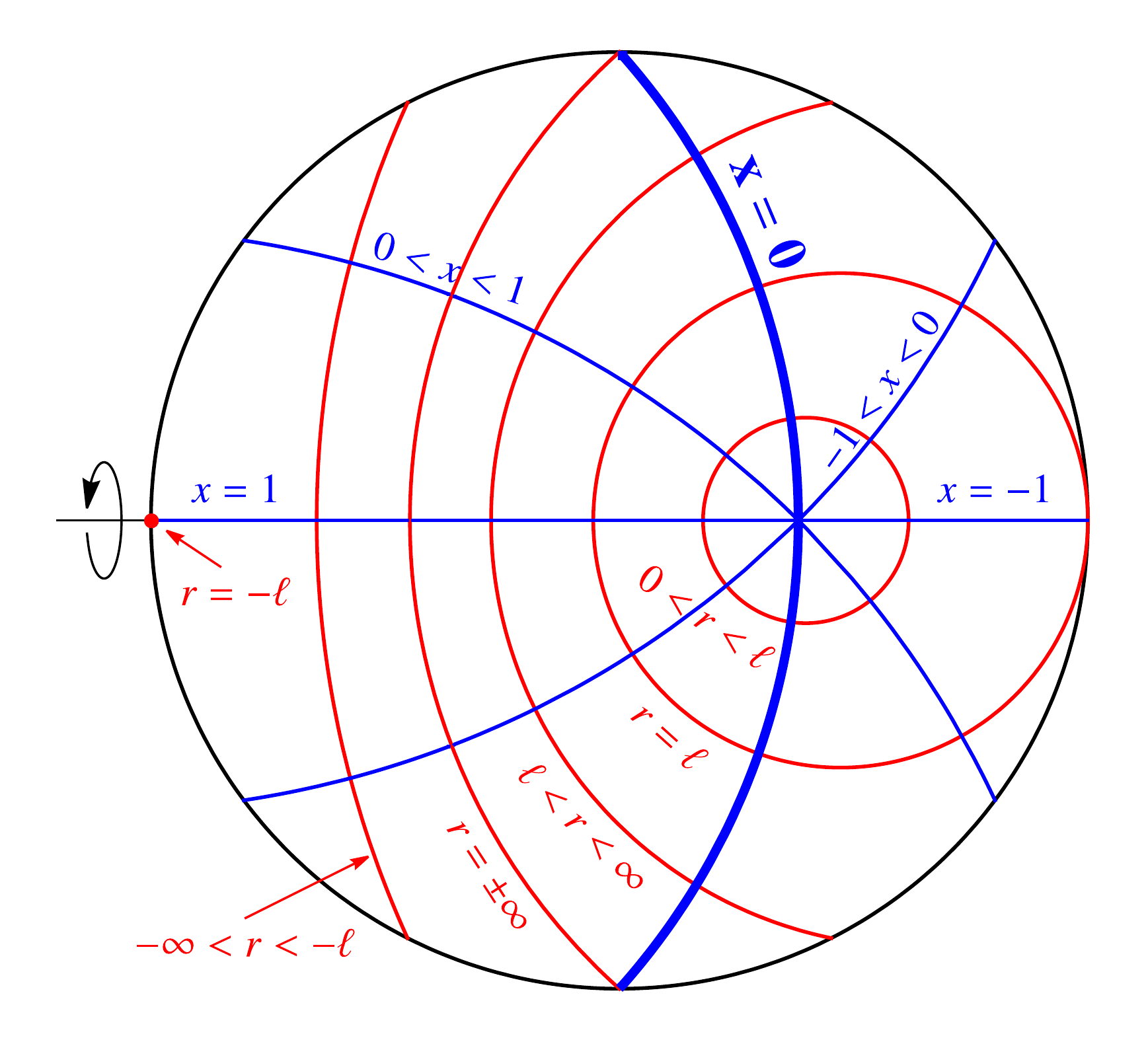}
  \includegraphics[width=.49\textwidth]{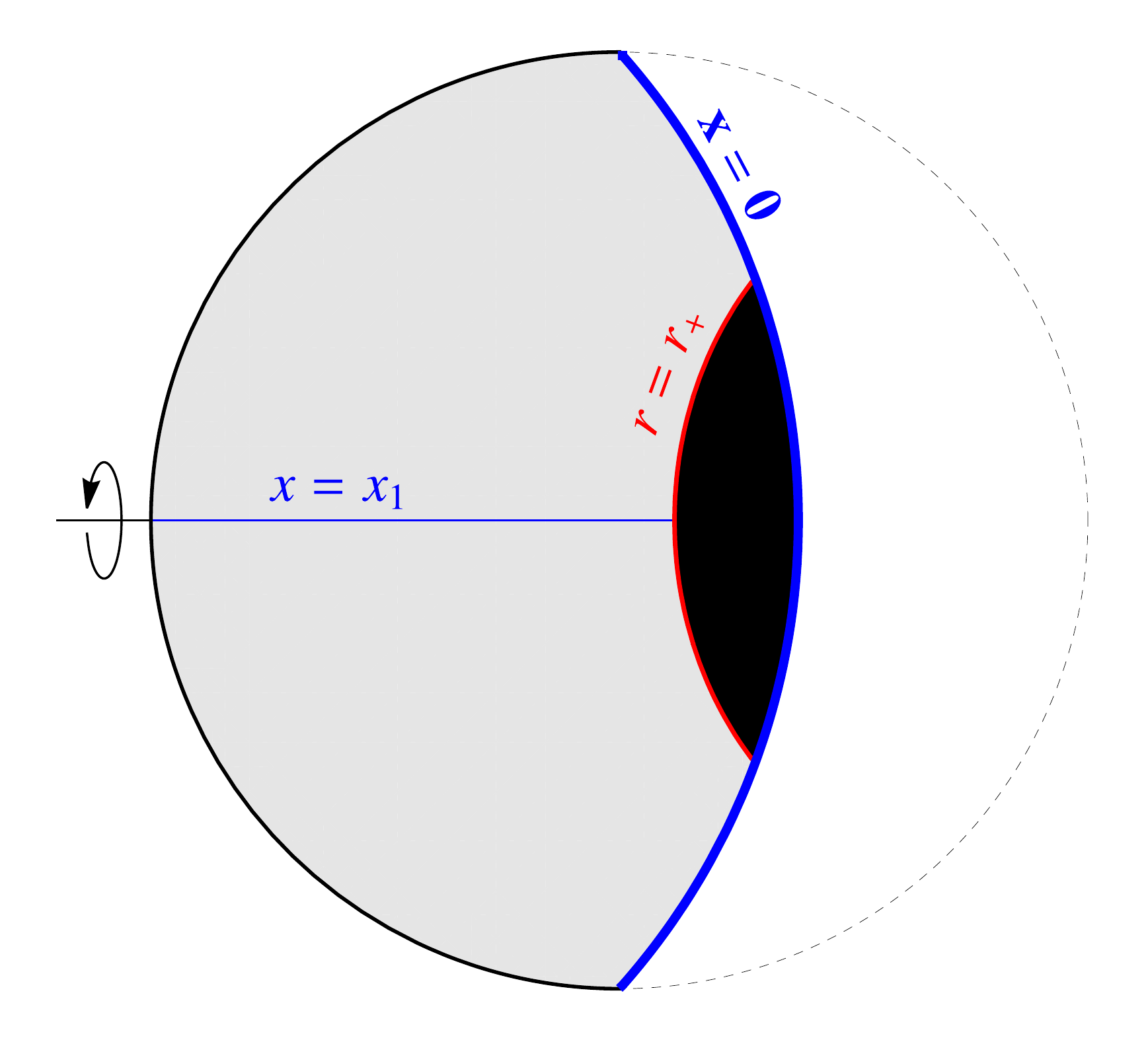}
\caption{\small Bulk geometry in a slice at constant $t$ and $\phi$. Left: C-metric coordinates $(x,r)$ in the spatial Poincar\'e disk of empty global AdS$_4$ ($\mu=0$, $\kappa=+1$). Lines of constant $x\in [-1,1]$ are blue arcs; lines of constant $r\in [-\infty,-\ell]\cup [0,\infty]$ are red arcs (full circles for $0<r\leq \ell$). The asymptotic boundary (black circle) is at $xr=-\ell$. The $\phi$ axis of rotation is $x=\pm 1$. Right: Sketch of braneworld construction with a black hole in it. The bulk is cut off at a brane at $x=0$ and only the (gray) region $0\leq x\leq x_1$ is retained; the root $x=x_1$ of $G(x)$ is now the $\phi$ axis. A second copy of this region, not shown, is glued at the brane to make a $\Z_2$-symmetric two-sided braneworld. A bulk black hole with event horizon at $r=r_+$ is attached to the brane. Dual three-dimensional fields satisfy transparent boundary conditions at the junction between the dynamical brane (thick blue) and the non-dynamical AdS$_4$ boundary (black).}
\label{fig:sketch}
\end{figure}

When $\mu\neq 0$ the geometry is more complicated, but the C-metrics have the nice feature that the surface $x=0$ is always totally umbilic.  That is, that the extrinsic curvature $K_{ab}$ and induced metric $h_{ab}$ satisfy
\beq\label{Kab}
K_{ab}= -\frac1{\ell} h_{ab}\,.
\eeq
The $x=0$ surface is therefore where we put the brane (figure~\ref{fig:sketch}). We cut the bulk geometry at $x=0$ and keep a range of positive $x$ to be specified later. It is easiest to understand the properties of this brane when $\mu=0$, since then it corresponds to the surface $\sigma=\sigma_b$ in \eqref{emptyads4} with
\beq\label{sigmanu}
\cosh\sigma_b=\frac{\ell_3}{\ell_4}=\sqrt{1+\frac{\ell_3^2}{\ell^2}}\,.
\eeq
The brane geometry is AdS$_3$ with curvature radius $\ell_3$. We take the brane to be two-sided, with $\Z_2$ orbifold boundary conditions on it (our main results extend to one-sided branes changing only factors of $2$). As a result, the metric is continuous across the brane, but its derivative is discontinuous on account of the brane stress tensor, which is
\beq
S_{ab}=-\frac{2}{8\pi G_4}(K_{ab}-h_{ab} K)=-\frac{1}{2\pi G_4 \ell}h_{ab}\,.
\eeq
From here, we see that the brane tension is
\beq\label{tension}
\tau=\frac{1}{2\pi G_4 \ell}\,.
\eeq
In the limit $\ell\to\infty$, with $\ell_3\to\ell_4$, the brane becomes tensionless: this is an equatorial section, $\sigma=0$, of the AdS$_4$ bulk. In the opposite limit $\ell\to 0$, the brane is located increasingly closer to the AdS$_4$ asymptotic boundary at $\sigma\to\infty$.\footnote{Since in this limit $\ell_4\to \ell$, we must rescale the entire metric by $\ell^2$ in order to keep its size finite. We return to this point below, and more specifically in appendix~\ref{app:zerobr}.}

Karch and Randall showed that when the brane tension is non-zero and finite ($0<\ell<\infty$ in our case), there is a massive graviton bound state localized on the brane \cite{Karch:2000ct}. That is, gravity on the brane is not of the ordinary kind, but a massive gravity theory. However, when $\ell$ is very small and the brane is very close to the boundary, the graviton is almost massless. This will be the regime of most interest, but our bulk construction applies to all values of $\ell$.

The effective three-dimensional theory can be obtained by solving the bulk Einstein equations in the `excluded region' between the AdS$_4$ boundary and the location of the brane, in an expansion for small $\ell$; holographically, this means integrating out the ultraviolet CFT degrees of freedom down to the cutoff energy $1/\ell$, which induces the gravitational dynamics on the brane.
Ref.~\cite{deHaro:2000wj} derived  in this way the effective field equations, but it is simpler to obtain the three-dimensional effective action using the results in \cite{Emparan:1999pm} (see also \cite{Chen:2020uac}). The result is
\beq\label{effact0}
I=\frac{\ell_4}{8\pi G_4}\int d^3 x \sqrt{-h}\left[ \frac{4}{\ell_4^2}\lp 1-\frac{\ell_4}{\ell}\rp+R+\ell_4^2\lp \frac{3}{8}R^2-R_{ab} R^{ab}\rp +\dots \right]+I_{\rm{CFT}}\,,
\eeq
where the three-dimensional metric and curvatures are those induced on the brane, and $I_{\rm{CFT}}$ is holographically defined by the bulk. The quadratic curvature terms are the same as in the `new massive gravity' theory of \cite{Bergshoeff:2009hq}, a feature that deserves further attention but which we will not pursue here. The dots indicate additional higher-curvature terms, which are proportional to increasing powers of $\ell_4$, or equivalently, powers of $\ell$ when $\ell$ is much smaller than $\ell_3$.

We now write this action as
\beq\label{effact}
I=\frac{1}{16\pi G_3}\int d^3 x \sqrt{-h}\left[ \frac{2}{L_3^2} +R+\ell^2\lp \frac{3}{8}R^2-R_{ab} R^{ab}\rp +\dots \right]+I_{\rm{CFT}}\,,
\eeq
where we have identified the effective three-dimensional Newton's constant as
\beq\label{G34}
G_3=\frac1{2\ell_4}G_4\,,
\eeq
and the three-dimensional cosmological constant term as
\beqa\label{L3ell3}
\frac{1}{L_3^2}=\frac{2}{\ell_4^2}\lp 1-\frac{\ell_4}{\ell}\rp=\frac{1}{\ell_3^2}\lp 1+\frac{\ell^2}{4\ell_3^2}\rp\,.
\eeqa
In the last step we have used \eqref{l34} expanded to quadratic order in $\ell$. The difference between $L_3$ and the curvature radius of the brane $\ell_3$ is that the latter receives a contribution from the higher-curvature terms.

Let us now see what the parameter $\ell$ corresponds to in the three-dimensional theory.
The number of microscopic degrees of freedom of the holographic dual CFT is measured by what, with a slight abuse of terminology, we will call the ``central charge'' $c$, and which for convenience we choose to normalize as
\beq\label{central}
c=\frac{\ell_4^2}{G_4}\,.
\eeq
If the CFT is an ABJM theory with parameters $N$ and $k$ (for the rank of the gauge group and the Chern-Simons level), then \cite{Aharony:2008ug}
\beq
c=3\frac{(2Nk)^{3/2}}{k}\,,
\eeq
but we will not need to be this specific about the CFT.

We can now use \eqref{central} together with \eqref{l34} to express the parameter $\ell$ in terms of magnitudes of the dual 3D theory,
\beq\label{nu3d}
\frac{\ell}{1+(\ell/\ell_3)^2}=2\,c\, G_3\,.
\eeq
This is an exact relation. We can write it in terms of $L_3$ in a perturbative expansion using \eqref{L3ell3}.
In the limit where $\ell$ is small,
\beq\label{smallnu3d}
\ell=  2\, c\,G_3\lp 1+\ord{ c\,G_3/L_3}^2\rp\,.
\eeq

We see that if we keep $c$ and $L_3$ finite, then when $\ell\to 0$ and the brane approaches the AdS$_4$ boundary, the gravitational coupling $G_3$ must vanish too. Gravity on the brane becomes weaker, and therefore when $\ell\to 0$ there is no backreaction of the CFT. In this model, the gravitational strength on the brane and the graviton mass are controlled by the same parameter ($\sigma_b$, or $\ell/\ell_3$). One may separate them by introducing an explicit Einstein-Hilbert term on the brane {\`a} la DGP \cite{Dvali:2000hr}, but in that case there is no known solution with a black hole localized on the brane.

At the opposite end, when $\ell\to\infty$ the graviton mass becomes as large as possible, and gravity on the brane is completely four-dimensional; the tensionless brane has no effect on bulk propagation, other than imposing a $\Z_2$ projection.

With \eqref{smallnu3d}, the effective action \eqref{effact} can be written in terms of only the parameters $G_3$, $L_3$ and $c$ of the three-dimensional theory. Higher order corrections in the action are suppressed when the curvature scales on the brane, typically $\sim L_3$, are longer than $\ell\simeq \ell_4\simeq c\, G_3$. Therefore $\ell$ is the cutoff length scale of the effective theory. 

On the other hand, the three-dimensional Planck length (temporarily restoring  $\hbar$) is
\beq\label{LPl3}
L_\text{Planck}^{(3)}=\hbar\, G_3\,.
\eeq
From \eqref{smallnu3d} we have
\beq\label{lggLpl}
\ell \sim c\, \hbar\, G_3\sim c\, L_\text{Planck}^{(3)}\gg L_\text{Planck}^{(3)}\,,
\eeq
since in holography the classical bulk requires large central charges, $c\gg 1$. The reason that the cutoff length $\ell$ is much larger than the three-dimensional Planck length is that the huge number $\sim c$ of CFT degrees of freedom enhance their contribution to loop corrections \cite{Dvali:2007hz}.

There is no contradiction between requiring $c\gg 1$ and performing a small $\ell$ expansion. The latter is actually an expansion in 
\beq\label{smallell}
\frac{\ell}{\ell_3}\sim \frac{c\, \hbar\, G_3}{L_3}\ll 1\,,
\eeq
as is required for the validity of the effective three-dimensional description, while the limit of large central charge is
\beq\label{largec}
c\sim \frac{\ell}{\hbar\, G_3} \gg 1\,,
\eeq
which, using \eqref{G34}, is equivalent to the semiclassical bulk limit $L_\text{Planck}^{(4)}/\ell_4\ll 1$. So both limits \eqref{smallell} and \eqref{largec} are simultaneously consistent if we consider solutions with $\ell\ll \ell_3$ and we neglect quantum bulk corrections.

With the understanding of \eqref{smallell}, we see that the leading CFT contributions, which  enter as proportional to $c$ in the action $I_{\rm{CFT}}$ in \eqref{effact}, are of linear order in $\ell/\ell_3$. This distinguishes them from higher-curvature corrections, which are $\propto (\ell/\ell_3)^2$ or higher.

\subsection{Global aspects of the bulk}

We can get a quick idea of how the bulk black hole appears localized on the brane by taking the tensionless limit $\ell\to\infty$. Rescaling
\beq\label{toSchAdS}
\mu=\frac{2m}{\ell}\,,\qquad r=\frac{\ell_3}{\ell}\rho
\eeq
and keeping $m$, $\rho$ and $\ell_4$ finite, the bulk metric \eqref{statc} becomes
\beq
ds^2= -\lp\frac{\rho^2}{\ell_4^2}+\kappa-\frac{2 m}{\rho}\rp dt^2+\frac{d\rho^2}{\frac{\rho^2}{\ell_4^2}+\kappa-\frac{2 m}{\rho}} +\rho^2\lp \frac{dx^2}{1-\kappa x^2}+ \lp 1-\kappa x^2\rp d\phi^2\rp\,,
\eeq
which is the metric of AdS$_4$ black holes. The brane at $x=0$ slices them through an extremal section (an `equator') of zero extrinsic curvature.

At the opposite end, when $\ell\to 0$ the solution becomes
\beq\label{zerobckr}
ds^2\to \frac{\ell^2}{r^2 x^2}\lp -\lp \frac{r^2}{\ell_3^2}+\kappa\rp dt^2 +\frac{dr^2}{\frac{r^2}{\ell_3^2}+\kappa}+r^2\lp \frac{dx^2}{G(x)}+G(x)d\phi^2\rp\rp\,.
\eeq
The brane is now at the asymptotic boundary  of AdS$_4$ at $x\to 0$. When $\kappa=-1$ the boundary geometry, where $G(0)=1$, has a black hole (BTZ, as we will see) on it with horizon at $r=\ell_3$. This limiting solution is indeed the same as the construction in \cite{Hubeny:2009rc}, which is based on the fact that \eqref{zerobckr} is a double Wick rotation of the Schwarzschild-AdS$_4$ solution, as we explain in appendix~\ref{app:zerobr}.

More generally, whether the solutions \eqref{statc} have black holes, and of what kind, depends on the character of the roots of $H$ and $G$. For instance, it is apparent from \eqref{statc} that the roots of $H(r)$ are Killing horizons of $\partial/\partial t$. We want that a positive root $r_+$ exists for the black hole horizon, but also that there are no acceleration (non-compact) horizons. In the $(x,\phi)$ sector, the real roots of $G(x)$ are symmetry axes (fixed-point sets) of $\partial_\phi$, and their properties determine the geometry and topology of the horizons. Since $H$ and $G$ are cubic functions, the analysis of these roots tends to be involved, but below we will deal with this by switching to more efficient parametrizations.

The question of regularity at symmetry axes of $\partial_\phi$ is a central point in the analysis of all the C-metrics. As argued in \cite{Emparan:1999fd}, in order to have a finite black hole in the bulk (instead of, say, an infinite black string) we must be in a regime of parameters where there is at least one positive root of $G(x)$, the smallest of which we will call $x_1$. Then we restrict $x$ to the range
\beq\label{xrange}
0\leq x\leq x_1\,.
\eeq
It is now much more convenient to use $x_1$ as a primary parameter, and consider $\mu$ as a derived one, with
\beq\label{mux1}
\mu =\frac{1-\kappa x_1^2}{x_1^3}\,.
\eeq
We get the desired parameter range by, first, assuming that $x_1>0$, and furthermore taking
\beq
x_1\in (0,1]\quad\text{for $\kappa=+1$}\,,
\eeq
and
\beq
x_1\in (0,\infty)\quad\text{for $\kappa=-1,0$}\,.
\eeq
Note that $\mu$ is a monotonically decreasing function of $x_1$, with $\mu\to\infty$ when $x_1\to 0$, and $\mu\to 0$ at the upper limit of $x_1$.

In order to avoid a conical singularity at $x=x_1$ we must identify
\beq
\phi \sim \phi +2\pi \Delta
\eeq
with
\beq\label{Delx1}
\Delta =\frac{2}{|G'(x_1)|}=\frac{2x_1}{3-\kappa x_1^2}\,.
\eeq
Then, sections of constant $t$ and $r$  with $x$ varying in \eqref{xrange} are topological disks with $x$ playing the role of radial coordinate. We may more conveniently think of them as caps, and $x$ as roughly equivalent to the cosine of the polar angle along the cap (see fig.~\ref{fig:sketch}). In a two-sided brane, we glue two of these caps along their rim at $x=0$, to form a lens shape.
Notice that  $G'(x_1)<0$ in the range of $x_1$ considered, and that $\Delta$ is independent of $\ell$.

\subsection{quBTZ: metric, mass, and stress tensor}

The metric induced on the brane at $x=0$ is
\beq
ds^2=-H(r)dt^2+\frac{dr^2}{H(r)}+r^2d\phi^2
\eeq
with $H(r)$ given in \eqref{HG}.  This metric is asymptotic to AdS$_3$ at $r\to\infty$, but the coordinates are not canonically normalized since $\phi$ does not have periodicity $2\pi$ but $2\pi\Delta$. In order to fix this, we rescale
\beq\label{bartphir}
t=\Delta\,\bar{t}\,,\qquad \phi=\Delta\,\bar\phi\,,\qquad r=\frac{\bar{r}}{\Delta}\,,
\eeq
so that now $\bar{\phi}\sim \bar{\phi} +2\pi$ and the metric takes the form
\beq\label{quBTZ}
ds^2=-\lp \frac{\bar{r}^2}{\ell_3^2}-8\mathcal{G}_3 M- \frac{\ell F(M)}{\bar{r}}\rp d\bar{t}^2
+\frac{d\bar{r}^2}{\frac{\bar{r}^2}{\ell_3^2}-8\mathcal{G}_3 M-\frac{\ell F(M)}{\bar{r}}}
+\bar{r}^2d\bar{\phi}^2\,,
\eeq
where $\mathcal G_3$ is a `renormalized Newton's constant' which we will soon describe, and $F(M)$ is a function we give later in \eqref{FM}.

The identification of the three-dimensional mass $M$ here, which is given by
\beq\label{Mstat}
M=-\frac{\kappa}{8G_3}\frac{\ell}{\ell_4}\Delta^2 =-\frac{1}{2G_3}\frac{\ell}{\ell_4}\frac{\kappa x_1^2}{(3-\kappa x_1^2)^2}\,,
\eeq
requires some care. In Einstein-AdS gravity the mass is obtained by identifying the subleading, constant term in $g_{\bar{t}\bar{t}}$ as $8G_3 M$. However, the higher curvature terms in the effective theory \eqref{effact} modify this definition by adding corrections beginning at order $\ell^2$. These can be computed from, e.g.~\cite{Cremonini:2009ih}, with the result that the correct mass is obtained from the same term in $g_{\bar{t}\bar{t}}$ but using a `renormalized' Newton's constant,
\beqa
\mathcal{G}_3&=&\lp 1-\frac{\ell^2}{2L_3^2}+\ord{\frac{\ell}{L_3}}^4\rp G_3\nn\\
&=&\frac{1-\frac{\ell^2}{2\ell_3^2}}{2\ell_4} G_4+\ord{\frac{\ell}{L_3}}^4
\,.\label{approxG}
\eeqa
But instead of this expansion, we will  use 
\beq\label{exactG}
\mathcal{G}_3=\frac{\ell_4}{\ell}G_3= \frac1{2\ell}G_4\,,
\eeq
which is equivalent to \eqref{approxG} up to at least $\ord{\ell/L_3}^4$.

We will work under the assumption that \eqref{Mstat} and \eqref{exactG} are valid to \emph{all} orders in $\ell$: our analysis of the first law will strongly point to this conclusion. In further support of this,  \cite{Emparan:1999fd} derived \eqref{exactG} exactly by integrating the bulk volume in the action with a natural infrared bulk cutoff at $r\to\infty$. The justification for this procedure here, however, is unclear, and we interpret \eqref{exactG} very differently: as a remarkably simple, all-order resummation of the higher-curvature corrections to the mass---since what we are correcting is how the constant $\mathcal{G}_3 M$ that appears in the metric relates to the physical $G_3 M$.

It should be possible to work out successive higher-curvature corrections to the effective action \eqref{effact}, and then to the mass, and verify that \eqref{exactG} correctly reproduces all these corrections, but this approach very quickly becomes extremely cumbersome. In the final section we discuss another possible, more direct manner of deriving \eqref{exactG}.

In our subsequent discussion, it will often be more convenient to fix the `renormalized' quantities $\mathcal{G}_3$ and $\ell_3$ that appear in the exact metric, rather than the `bare' parameters in the effective action, $G_3$ and $L_3$.

We see in \eqref{Mstat} that $\mu$, or $x_1$, is a parameter that controls the mass of the solution. We have also introduced
\beq\label{FM}
F(M)=\mu \Delta^3=8\frac{1-\kappa x_1^2}{(3-\kappa x_1^2)^3}\,,
\eeq
which depends on $\mathcal{G}_3 M$ through $x_1$, but is otherwise independent of $\ell/\ell_3$ so it does not change as we vary the strength of the backreaction for fixed $\mathcal{G}_3 M$. Actually, all physical magnitudes depend on $\kappa x_1^2$ and not on $\kappa$ or $x_1$ separately.

According to the holographic dictionary, the metric induced on the brane solves the semiclassical equations \eqref{holobackr}. We can therefore identify the holographic CFT stress tensor as the right-hand side of the gravitational equations derived from the action \eqref{effact}, which are
\beqa\label{3Deffeqn}
8\pi G_ 3\langle T_{ab} \rangle&=&R_{ab}-\frac12 h_{ab} \lp R+\frac2{L_3^2}\rp\nn\\
&&+\ell^2\Biggl[  4 R_a{}^{c}R_{bc}-\frac94 R R_{ab}-\nabla^2 R_{ab} +\frac14 \nabla_a \nabla_b R\nn\\&&\qquad+\frac12 h_{ab} \lp \frac{13}8 R^2- 3R_{cd}R^{cd}+\frac12 \nabla^2 R\rp \Biggr] +\dots
\eeqa
We decompose this stress tensor as
\beq\label{totTab}
\langle T^a{}_b \rangle=  \langle T^a{}_b \rangle_0+\ell^2\langle T^a{}_b \rangle_2+\dots
\eeq
where
\beq\label{Tab0}
8\pi G_ 3\langle T^a{}_b \rangle_0= R^a{}_b-\frac12\delta^a{}_b \lp R+\frac2{\ell_3^2}\rp\,,
\eeq
and
\beqa\label{Tab2}
8\pi G_ 3\langle T^a{}_b \rangle_2&=&4 R^{ac}R_{bc}-\frac94 R R^a{}_b-\nabla^2 R^a{}_b +\frac14 \nabla^a \nabla_b R\nn\\ &&+\frac12\delta^a{}_b \lp \frac{13}8 R^2- 3R_{cd}R^{cd}+\frac12 \nabla^2 R-\frac1{2\ell_3^4} \rp
\,.
\eeqa
Note that in $\langle T^a{}_b \rangle_0$ we have absorbed the $\ord{\ell^2}$ difference between $L_3$ and $\ell_3$, which is compensated by adding a constant to $\langle T^a{}_b \rangle_2$.  By doing this, each term in the expansion \eqref{totTab} separately vanishes in the ground state for arbitrary $\ell$ (they are also separately conserved).

Computing these terms for the exact metric \eqref{quBTZ} in $(\bar{t},\bar{r},\bar{\phi})$ coordinates we find
\beq\label{renst}
\langle T^a{}_b \rangle_0=\frac{\ell}{16\pi G_3} \frac{F(M)}{\bar{r}^3}\,\text{diag}\{1,1,-2\}
\eeq
and
\begin{align}
\langle T^a{}_b \rangle_2=\frac{\ell}{16\pi G_3} \frac{F(M)}{\bar{r}^3} \biggl(& \frac1{2\ell_3^2}\,\text{diag}\{1,-11,10\} -\frac{24 \mathcal{G}_3 M}{\bar{r}^2} \,\text{diag}\{3,1,-4\}\nn\\ &+\frac{\ell F(M)}{2\bar{r}^3} \,\text{diag}\{-29,-17,43,\} \biggr)\,.
\end{align}
The last term gives a non-zero trace,
\beq
\langle T^a{}_a \rangle_2=\frac1{16\pi G_3}\lp \frac38 R^2 - R_{ab}R^{ab}-\frac3{2\ell_3^4}\rp=-\frac{3\ell^2}{32\pi G_3}\frac{F(M)^2}{\bar{r}^6}\,.
\eeq
The reason is that the conformal symmetry is broken in the effective theory by the cutoff $\ell$.

We can now use \eqref{nu3d} to eliminate $\ell$ and express the results solely in terms of 3D magnitudes. This is simpler for small $\ell$,  \eqref{smallnu3d}, where we find
\beq\label{renst2}
\langle T^a{}_b \rangle=\frac{c}{8\pi} \frac{F(M)}{\bar{r}^3}\,\text{diag}\{1,1,-2\}\lp 1+\ord{ c\,G_3/\ell_3}^2\rp\,.
\eeq

Since the backreaction vanishes when $\ell\to 0$, the metric \eqref{quBTZ} in this limit is interpreted as a classical solution, and indeed these geometries solve the classical Einstein-AdS equations and are locally AdS$_3$. In this limit of small backreaction, the corrections to the classical geometry are proportional to $c\,G_3$, as expected. However, we emphasize that the CFT is consistently solved simultaneously with the 3D gravitational equations, and therefore yields the exact backreaction of the light CFT degrees of freedom for finite $\ell$, and also of heavy ones through the non-perturbative resummation of all higher-curvature corrections.

Since we recover the classical BTZ black hole for $\ell=0$ and $\kappa=-1$, when $\ell>0$ we refer to \eqref{quBTZ} as the quantum BTZ black hole, or quBTZ. This interpretation was initially given in \cite{Emparan:2002px}. In contrast with the classical BTZ solution, whose curvature is constant, the curvature of quBTZ varies and in fact blows up at a singularity at $\bar{r}=0$, a feature that is inherited from the curvature singularity that \eqref{statc} has at $r=0$.

Interestingly, since $\ell/\ell_3$ does not enter directly in $F(M)$, it depends on backreaction only through the rescaling between $G_3$ and $\mathcal{G}_3$. Then, the dependence of the stress tensor on the black hole mass $M$ at low orders in $\ell$ is simple. There is no term in $\langle T^a{}_b \rangle$ that is $\propto \ell^2$, and all the subsequent dependence on $\ell$, starting at $\ell^3$, is due to the higher-curvature corrections. This intriguing result, we will see, extends to rotating solutions.

\subsection{Branches of holographic quantum black holes}\label{subsec:branches}

An important feature of the holographic construction, already observed in \cite{Emparan:1999fd, Emparan:2002px}, is that the range of masses covered by \eqref{Mstat} is finite,
\beq\label{massrange}
-\frac1{8\mathcal{G}_3}\leq M\leq \frac1{24 \mathcal{G}_3}\,.
\eeq
In the following analysis the only dependence on $\ell$ enters through rescaling from $\mathcal{G}_3$ to $G_3$. We will then consider that we work with fixed $\mathcal{G}_3$.

Negative masses are obtained when $\kappa=+1$, with the minimum reached for $x_1=1$ ($\mu=0$), and zero mass for $x_1\to 0$ ($\mu\to\infty$). We refer to these solutions as
\beq\label{b1a}
\text{Branch $1a$:}\qquad \kappa=+1,\quad 0<x_1<1\,.
\eeq

In the range of positive masses, with $\kappa=-1$, we can have two solutions with the same $M$ but different values of $x_1$. Then we have two branches of solutions over the same mass range, which we denote as
\beqa\label{b1b}
&\text{Branch $1b$:}\quad &\kappa=-1,\quad 0<x_1<\sqrt{3}\,,\label{b1b}\\
&\text{Branch 2:}\quad &\kappa=-1,\quad  \sqrt{3} <x_1<\infty\,.
\eeqa

The two branches meet at the upper mass bound where $x_1=\sqrt{3}$. For $M=0$ we have two distinct solutions: one of them in branch 1 with $x_1\to\infty$ ($\mu\to 0$), and the other in branch 2 with $x_1=0$ ($\mu=\infty$). Since physical magnitudes depend only on the combination $\kappa x_1^2$ the branches $1a$ and $1b$ join smoothly at $x_1= 0$, and we may characterize the branches more concisely as
\beqa
\text{Branch 1:}&\quad  -1&<-\kappa x_1^2<3\,,\\
\text{Branch 2:}&\quad 3 &<-\kappa x_1^2<\infty\,.
\eeqa
The value $x_1=0$ in branch 1 is the same as the $\kappa=0$ solutions, so we shall not discuss this last case separately.

We also add a Branch 3 of `BTZ black strings'. For all $M\geq 0$ we can have an exact bulk solution which is the geometry \eqref{emptyads4} with $\kappa=-1$ and with the appropriate periodic identification of $\phi$ to yield a mass $M$ on the brane. It is a black string since each section at constant $\sigma$ has a BTZ black hole in it.\footnote{Following \cite{Emparan:1999fd} we assume an infrared bulk cutoff at $r\to\infty$.} Then, this exact solution gives an uncorrected BTZ black hole on the brane, and vanishing CFT stress energy tensor.

\begin{figure}
\centering
  \includegraphics[width=.8\textwidth]{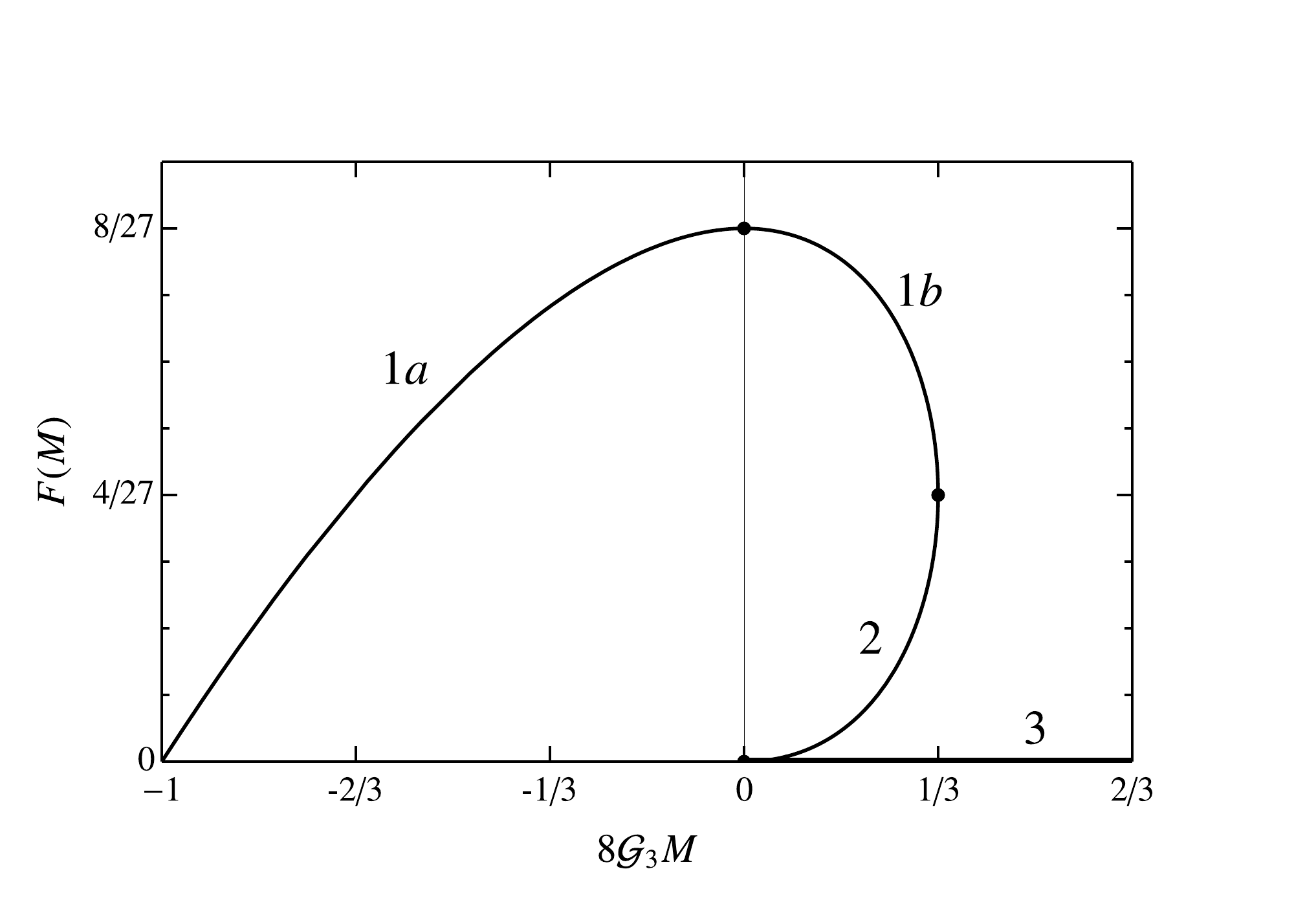}
\caption{\small The holographic stress-energy function $F(M)$ \eqref{FM} for the three branches $1a$, $1b$ and 2, of quantum black hole solutions. We also include a branch $3$ of bulk BTZ black strings, which give BTZ on the brane with $F=0$ for all $M\geq 0$. }
\label{fig:FofM}
\end{figure}

In fig.~\ref{fig:FofM} we plot the function $F(M)$ for all branches. Our discussion here and in the next subsection will loosely follow \cite{Emparan:2002px} (to which we refer for other details), with some relevant additions.

The lower mass limit $-1/(8\mathcal{G}_3)$ in branch $1a$ is the mass of global AdS$_3$, reached for $\kappa=+1$, $x_1=1$, where the renormalized stress tensor vanishes. Above this ground state,  the solutions with negative masses correspond, in the limit $\ell\to 0$ of \eqref{quBTZ}, to conical singularities in AdS$_3$. The CFT stress-energy tensor in these horizonless backgrounds is a Casimir effect. When $\ell>0$ the geometries  are interpreted as quantum-corrected conical singularities. As first discussed in \cite{Emparan:2002px}, the backreaction of this quantum Casimir stress tensor dresses the singularity with a horizon, in a sort of `quantum cosmic censorship' (see also \cite{Casals:2016odj,Casals:2019jfo}). This quantum dressing is also present for conical singularities in branes that are asymptotically locally flat.

The upper mass bound in branches $1b$ and 2 is intriguing. One might want to attribute it to the peculiar infrared behavior of the massive gravity theory, but this cannot be the full explanation, since the bound is independent of $\ell/\ell_3$ and therefore applies also for $\ell\to 0$ where the 3D graviton mass vanishes and there is no gravitational dynamics. The bound seems to say that there is an upper limit, not on all quantum corrected BTZ black holes, but rather on those for which the CFT is captured by these localized bulk black holes. It is therefore not a feature of braneworld gravity, but a consequence of holographically representing the CFT by four-dimensional bulk gravity. Intuitively, the bulk black hole cannot extend into the bulk beyond the `throat' at the minimal AdS$_3$ slice of AdS$_4$, which puts an infrared limit on the validity of the effective 3D description. If this interpretation is correct, then an upper mass limit for non-trivial holographic quantum effects in AdS black holes is also expected in higher dimensions.

In the absence of any other known bulk solution, the only option for masses above the range \eqref{massrange} is branch 3: the BTZ black string. The holographic CFT in BTZ black holes with $M>1/(24 \mathcal{G}_3)$ is then in an unexcited state. This is a generic feature of the leading planar limit of holographic CFTs when the bulk is of `black string type'.

In the range $0<M<1/(24 \mathcal{G}_3)$, there are three bulk solutions (branch $1b$, 2 and 3) that correspond to different states of the CFT. Which one is preferred may depend on which has bigger quantum entropy. This was discussed extensively in \cite{Emparan:1999fd}, and will be more briefly revisited below.

It is natural to regard branch $1b$ as a continuation of branch $1a$ black holes. These, we have argued, must be considered as black holes formed from the backreaction of Casimir stress-energy. Then, we expect that the large stress-energy of the state $M=0$ in $1b$ should be dominated by Casimir energy. In contrast, branch 2 solutions start at $M=0$ with zero stress energy. In this case, it might perhaps be more appropriate to regard the stress tensor for $M>0$ as only due to the quantum Hawking radiation in equilibrium with a finite-temperature black hole, with (we speculate) the larger Casimir energy having been subtracted from the state.

\subsection{Validity of the solutions}

One might worry that the range of masses \eqref{massrange}, where $M\sim 1/G_3$, is always `Planckian' and hence quantum gravitational effects should be important and invalidate the semiclassical description. However, this is not the case. In contrast to four (or higher) dimensions, in three dimensions the relation $M\sim 1/G_3$ does not involve $\hbar$ and hence black holes with these masses need not be subject to large quantum fluctuations. Indeed, there is no such thing as a quantum Planck mass or energy in three dimensions. Instead one should refer to the quantum Planck length $L_\text{Planck}^{(3)}$ in \eqref{smallnu3d} (or to the Planck time). 

Cutoff effects will be small in our black holes if their horizon radius is $\gg \ell$. In this case, \eqref{lggLpl} implies that they will also be good semiclassical solutions, much larger than $L_\text{Planck}^{(3)}$.
It is easy to see from \eqref{quBTZ} that the typical radius of quBTZ black holes in branches 1$b$ and 2 is $\sim \ell_3$. Then, these are valid solutions in the regime $\ell_3\gg \ell$ in which the effective theory applies. The branch 1$a$ quantum-dressed cones have horizon radii $\sim (\ell_3^2 \ell)^{1/3}$. When $\ell_3\gg \ell$ this size is $\gg \ell$, and hence also safely within the regime of applicability of the effective theory.

\subsection{Comparing calculations of $\langle T^a{}_b \rangle$}

The renormalized stress-energy tensor for a free conformal scalar field in the BTZ black hole, and in negative-mass conical geometries, has been computed in \cite{Steif:1993zv,Shiraishi:2018pdw,Lifschytz:1993eb,Casals:2016odj,Casals:2019jfo}. The stress tensor depends on the boundary conditions for the quantum field at the asymptotic AdS$_3$ boundary. The holographic calculation naturally selects transparent conditions, where the fields propagate smoothly between the brane and the non-dynamical AdS$_4$ boundary (see fig.~\ref{fig:sketch}). In fact, the transparent boundary is responsible for the mass of the 3D graviton \cite{Karch:2000ct,Geng:2020qvw}. In the following we will only compare to free-field calculations made with these boundary conditions.

The renormalized $\langle T^a{}_b \rangle$ for the free conformal scalar takes the same form as \eqref{renst2}, where now, if we (arbitrarily) put $c=1$, the function $F(M)$ is given by
\beq\label{FMBTZ}
F(M)=\frac{(8G_3 M)^{3/2}}{2\sqrt{2}}\sum_{n=1}^\infty \frac{\cosh 2n\pi\sqrt{8G_3M}+3}{\lp\cosh 2n\pi\sqrt{8G_3M}-1\rp^{3/2}}
\eeq
for BTZ black holes with $M>0$, and by
\beq\label{FMns}
F(M)=\frac{(-8G_3 M)^{3/2}}{4\sqrt{2}}\sum_{n=1}^{N-1} \frac{\cos 2n\pi\sqrt{-8G_3M}+3}{\lp1-\cos 2n\pi\sqrt{-8G_3M}\rp^{3/2}}
\eeq
for conical singularities with $\sqrt{-8G_3M}=1/N$, where $N\in \Z^+$ \cite{Casals:2019jfo}. Note that for the BTZ black hole neither the radial dependence $\propto 1/r^3$ of the stress tensor, nor the precise tensorial structure $\text{diag}\{1,1,-2\}$ are uniquely preordained by conformal symmetry. A more complicated radial dependence, and other structures such as $\text{diag}\{-2,1,1\}$ (which is the form for a thermal plasma) or $\text{diag}\{ 1,0,-1\}$ are allowed, and indeed are present when asymptotic boundary conditions other than transparency are imposed \cite{Lifschytz:1993eb}.\footnote{The structure $\text{diag}\{1,0,-1\}$ appears in the Cotton tensor of \eqref{quBTZ}. This suggests that for non-transparent boundaries the effective gravitational action on the brane also includes Lorentz-Chern-Simons terms, yielding more general massive gravities \cite{Deser:1981wh,Bergshoeff:2009aq}.} Ref.~\cite{Lifschytz:1993eb} verified that the Green's function of the quantum field is in the Hartle-Hawking state and satisfies the KMS condition at the black hole temperature. Therefore, even if the stress tensor does not have the structure $\text{diag}\{-2,1,1\}$, it does have thermal character.

The sums in \eqref{FMBTZ} and \eqref{FMns} come from solving the free field theory using the method of images. It is interesting that at strong coupling, where this method would not be applicable, the holographic approach yields simpler, `resummed' expressions. When we add rotation the simplification will be even more dramatic.

\begin{figure}
\centering
  \includegraphics[width=.8\textwidth]{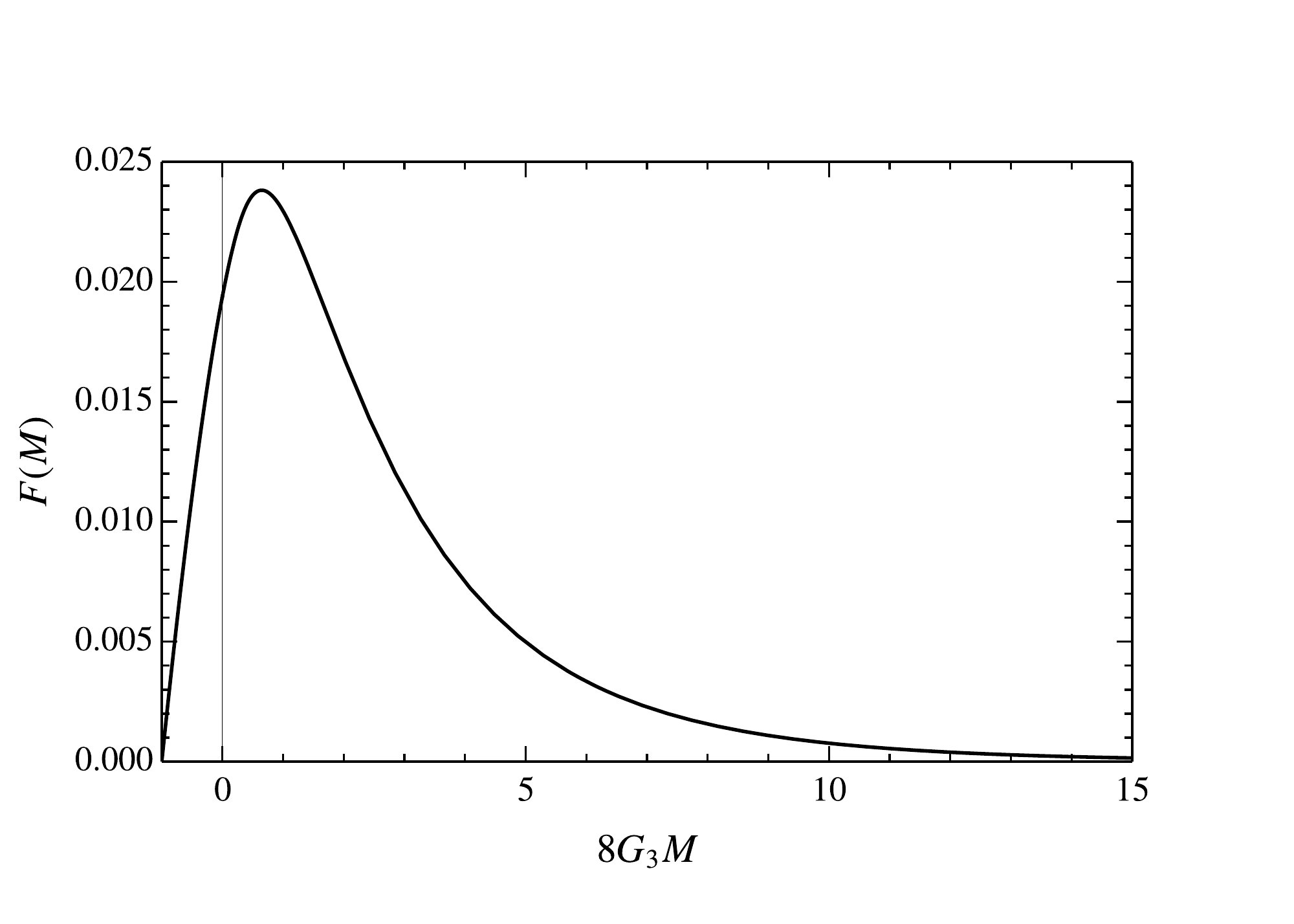}
\caption{\small The stress-energy function $F(M)$, \eqref{FMBTZ} and \eqref{FMns}, for a free conformal scalar.}
\label{fig:FofMfree}
\end{figure}

We plot $F(M)$ for free fields in fig.~\ref{fig:FofMfree} (for negative mass we interpolate between discrete values).
We see that the shape is qualitatively similar to fig.~\ref{fig:FofM} in the negative mass regime, and less so for positive masses. Quantitative comparisons cannot be made unambiguously since the number and type of fields, and their interactions, are very different in each case, but if we match the slopes of the curves at the AdS$_3$ vacuum, $8G_3M=-1$, then at larger mass the free field $F$ is greater than that of the holographic CFT. For free scalars the stress tensor extends smoothly to arbitrarily large $M$, and although it decreases with increasing $M$, it is never zero. Contrastingly, the stress tensor in the planar holographic calculation completely shuts off above $M=1/(24G_3)$.

The studies of backreaction of the free CFT have been limited to the perturbative regime of small linearized corrections. Since the stress tensor for the free and holographic CFTs has the same radial dependence and the same tensorial structure in both cases, differing only in $F(M)$, the metric corrections obtained in \cite{Lifschytz:1993eb,Shiraishi:2018pdw,Martinez:1996uv} have the same form as our solutions expanded to linear order in small $\ell$. They only differ in the mass dependence of the coefficients.

Finally, the holographic calculation of the renormalized stress tensor in the BTZ background in \cite{Hubeny:2009rc} yields \eqref{renst2} with the same $F(M)$, since we have shown that the construction in  \cite{Hubeny:2009rc} is the limit $\ell\to 0$ of the one in this paper.

\subsection{Quantum black hole horizons}

Now we turn to investigating the black hole event horizon. This lies at a positive real root $r=r_+$ of $H(r)$.
Then the circle radius of the horizon on the brane is $\bar{r}_+=\Delta\, r_+$.

We assume that $\ell/\ell_3$ and $\mu$ lie in ranges where such a root exists. Again, it will be more convenient to switch to another parametrization, based on the roots themselves, that deals with this automatically. Following \cite{Emparan:1999fd} we introduce the real, non-negative parameter
\beq\label{zdef}
z=\frac{\ell_3}{r_+ x_1}\,,
\eeq
and instead of $\ell$, a dimensionless parameter
\beq\label{nudef}
\nu=\frac{\ell}{\ell_3}\,.
\eeq
We can now eliminate $x_1$, or $\mu$, and $r_+$ using that
\beqa
x_1^2&=&-\frac1{\kappa}\frac{1-\nu z^3}{z^2(1+\nu z)}\,,\\
r_+^2&=&-\ell_3^2\kappa \frac{1+\nu z}{1-\nu z^3}\,,\\
\mu x_1 &=&-\kappa \frac{1+z^2}{1-\nu z^3}\,.
\eeqa
With $\nu$ and $z$ as parameters the mass is
\beq\label{Mznu}
M=\frac1{2\mathcal{G}_3}\frac{z^2(1-\nu z^3)(1+\nu z)}{(1+3z^2+2\nu z^3)^2}\,,
\eeq
with
\beq\label{cGnu}
\mathcal{G}_3=\frac{G_3}{\sqrt{1+\nu^2}}\,,
\eeq
and the coefficient of the stress tensor is
\beq
F(M)=8\frac{z^4(1+z^2)(1+\nu z)^2}{(1+3z^2+2\nu z^3)^3}\,.
\eeq
In this form it is not apparent that $F$ depends only on $\mathcal{G}_3 M$ and not separately on $\nu$, but of course it is still true since $\partial_\nu F-\partial_z F\, \partial_\nu (\mathcal{G}_3 M)/\partial_z (\mathcal{G}_3 M)=0$. In this parametrization $\kappa$ is not present in the expressions for physical quantities. It corresponds to
\beq
\kappa=\text{sign}(\nu z^3-1)\,,
\eeq
so we cover the entire range of branches 1 and 2 of bulk black holes of finite size by letting
\begin{equation}
0\leq \nu, z<\infty\,.
\end{equation}

\begin{figure}
\centering
  \includegraphics[width=.8\textwidth]{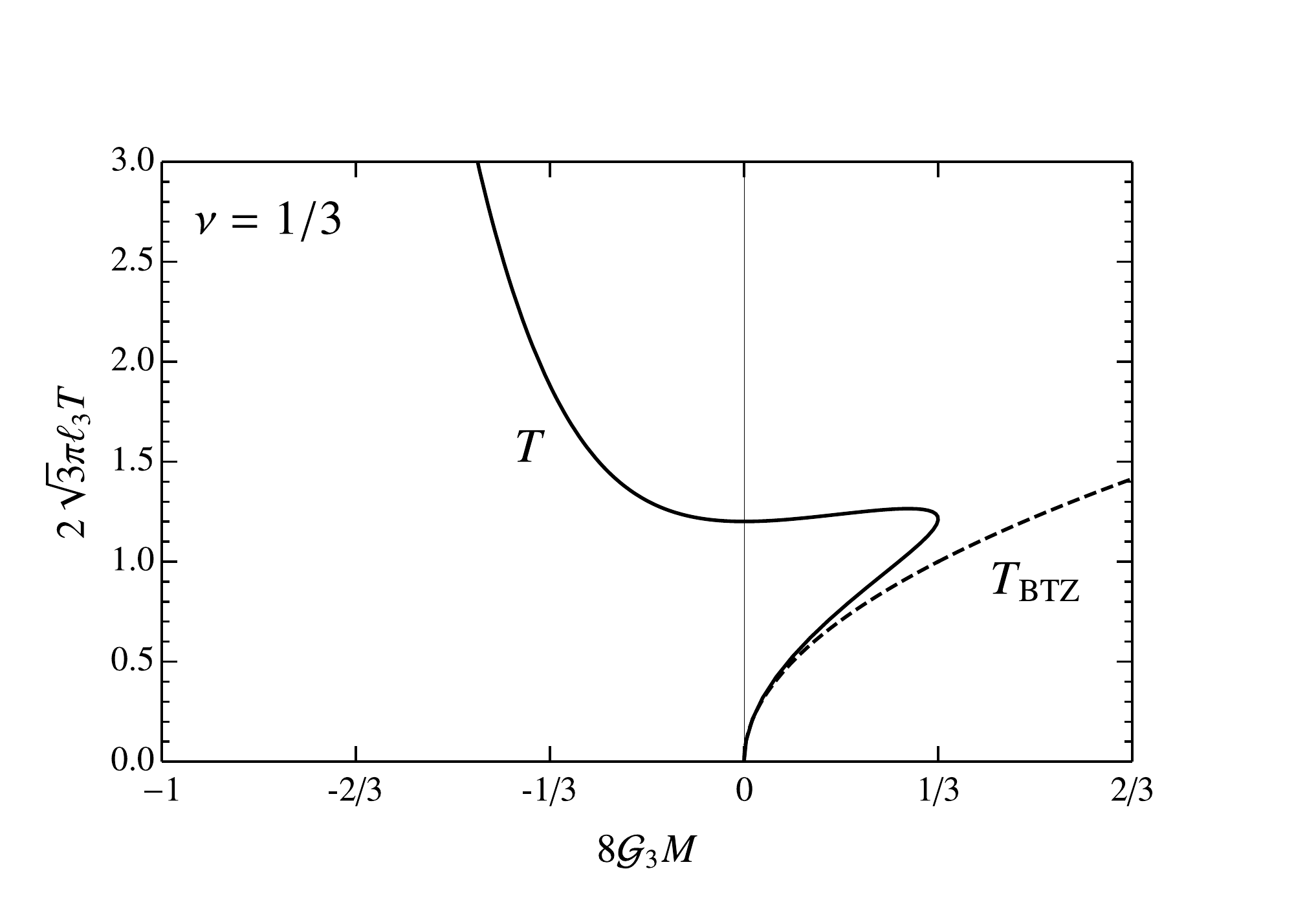}
\caption{\small Temperature of the quantum black holes and the classical BTZ black hole for given mass $M$.}
\label{fig:T}
\end{figure}

The temperature of the horizon, relative to the canonical timelike Killing vector on the brane, $\partial/\partial \bar{t}$, is
\beqa\label{Tznu}
T&=&\frac{\Delta H'(r_+)}{4\pi}\nn\\
&=&\frac1{2\pi \ell_3}\frac{z(2+3\nu z +\nu z^3)}{1+3z^2+2\nu z^3}\,.
\eeqa
The expressions are implicit but a plot (fig.~\ref{fig:T}) shows that the quantum black holes in both branches have higher temperature than the classical one with the same mass.
Observe that branch $1a$ black holes (with $M<0$ in fig.~\ref{fig:T}) have negative specific heat $\partial M/\partial T$. We interpret this in the same manner as in Schwarzschild or small AdS black holes: these black holes are too small and too hot to reach equilibrium with their Hawking radiation in the AdS$_3$ box, and will evaporate completely. For positive masses, there are two branches of quBTZ black holes, and branch $1b$ are the hotter of the two and branch 2 the colder. The specific heat of branch 2 is always positive, while that of branch $1b$ is more intricate.  We find that for any fixed $\nu$, the specific heat diverges for two black holes in branch $1b$: the $M=0$ black hole and another one at a certain finite mass $M_1(\nu)$ (involving an unilluminating cubic root). Since, generically, energy fluctuations in a thermal state are given by
\beq
\langle \delta E^2\rangle =T^2\frac{\partial E}{\partial T}\,,
\eeq
we expect that these black holes will be susceptible to large thermodynamic fluctuations and presumably be unstable.

Intriguingly, the largest black holes in branch $1b$, with masses in $M_1<M<1/(24 \mathcal{G}_3)$, also have negative specific heat. This could be an indication that these black holes can evaporate by radiating through the transparent interface to the non-dynamical region of the boundary, as in \cite{Almheiri:2019psf}. Perhaps this is also related to the large thermal fluctuations of the massless quBTZ in $1b$.

\subsection{Quantum entropy and the first law}

For the entropy associated to the bulk horizon, holographically interpreted as the quantum-corrected, generalized entropy, we obtain
\beqa\label{Squst}
S_\text{gen}&=&\frac{2}{4 G_4}\int_0^{2\pi \Delta}d\phi\int_0^{x_1}dx\, r_+^2\frac{\ell^2}{\lp \ell+ r_+ x\rp^2}\nn\\
&=& \frac{2\pi\ell_3^2}{G_4}\frac{\nu z}{1+3z^2+2\nu z^3}\nn\\
&=& \frac{\pi \ell_3}{G_3}\frac{z\sqrt{1+\nu^2}}{1+3z^2+2\nu z^3}\,.
\eeqa
In the last line we have converted to 3D units, so the result can be compared with the `classical' Bekenstein-Hawking entropy from the area of the horizon on the brane (which includes backreaction)
\beqa\label{Sclst}
S_\text{cl}&=&\frac1{4G_3}2\pi r_+\Delta\nn\\
&=& \frac{1+\nu z}{\sqrt{1+\nu^2}}S_\text{gen}\,.
\eeqa
In the theory \eqref{effact} with quadratic curvature terms, one should actually consider the Wald entropy. Following \cite{Jacobson:1993vj}, it is given by
\beq\label{SW}
S_W=\frac1{4G_3}\int dx \sqrt{q}\lp 1+\ell^2
\lp \frac34 R-g^{ab}_\perp R_{ab}\rp+\ord{\ell/\ell_3}^4\rp\,,
\eeq
where the integral is over the horizon with induced metric $q_{ab}$, $R$ and $R_{ab}$ are the three-dimensional spacetime curvatures evaluated on the horizon, and  $g^{ab}_\perp=g^{ab}-q^{ab}$ is the metric in directions orthogonal to the horizon. Evaluated on the quBTZ geometry \eqref{quBTZ}, we find  that
\beqa\label{Swald}
S_W&=&\lp 1-\ell^2\lp \frac1{2\ell_3^2}+\frac{\ell\mu}{r_+^3}\rp+\ord{\ell/\ell_3}^4\rp S_\text{cl}\\
&=&\lp 1-\frac{\nu^2}{2}-\nu^3\frac{z(1+z^2)}{1+\nu z}+\ord{\nu^4}\rp S_\text{cl}\,,
\eeqa

Properly, our results for effective three-dimensional magnitudes should always be expanded for small $\nu$, but it is sensible to consider $S_\text{gen}$ as exactly valid to all orders in $\nu$, since it is defined as a bulk magnitude that is exact up to quantum bulk corrections.

In the limit $\nu\to 0$ we recover the correct BTZ result in the absence of any backreaction,
\beq\label{limBTZ}
S_\text{gen},\, S_\text{cl},\, S_W\xrightarrow{\nu\to 0}\, S_\text{BTZ}=\frac{\pi^2\ell_3^2}{G_3}T=\pi\ell_3\sqrt{\frac{2 M}{G_3}}.
\eeq
For non-zero $\nu$ the three entropies $S_\text{cl}$, $S_W$ and $S_\text{gen}$ differ, but  the leading contributions to the entropy from the CFT must be proportional to $c$, or equivalently, since
\beq
c\simeq \nu \frac{\ell_3}{2G_3}\,,
\eeq
they must be linear in $\nu$. This distinguishes them from the higher-curvature corrections, which at their lowest order are $\propto \nu^2$. Neglecting the latter, the contribution to the quantum entropy from the entanglement entropy of the CFT fields outside the horizon is\footnote{The terms including up to $\nu^3$ here are reliably computed in our approximation, but we will not need them.}
\beq\label{Sout}
S_\text{out}=S_\text{gen}-S_W=-\nu z S_\text{BTZ}+\ord{\nu^2} \,.
\eeq
The fact that this is negative does not mean that the entanglement entropy of the CFT is negative. This $S_\text{out}$ corresponds to the finite part of the entanglement entropy after the leading piece has been absorbed in a renormalization of $G_3$,\footnote{Actually, since gravity on the brane is induced, the bare Newton constant is zero. Do not confuse this renormalization of $G_3$ with the higher-curvature effects on the mass that $\mathcal{G}_3$ accounts for in \eqref{approxG} and \eqref{exactG}.} so $S_\text{out}$ need not have a definite sign. The same negative sign is also present in branes that are asymptotic to Minkowski$_3$ \cite{Emparan:2006ni}.

The effect of the leading contribution to entanglement entropy can be gleaned from the change in the classical Bekenstein-Hawking entropy due to quantum backreaction effects,
\beq
S_\mathrm{ent}^{(0)} =S_\text{cl}(\nu,M)-S_\text{BTZ}(M)=\nu\frac{z(1+z^2)}{1+3z^2}S_\text{BTZ}(M)+\ord{\nu^2}\,.
\eeq
This quantity, linear in $\nu$, is finite, and moreover it is positive. It is a CFT effect on the Bekenstein-Hawking entropy different than the Wald corrections induced by higher-curvature terms.

In general $S_\text{out}$ will be dominated by the entanglement across the horizon in quantum states of the CFT with large Casimir effects. However, as we have argued above, for the $M=0$ BTZ black hole these Casimir effects appear to be small and we expect that instead thermal effects dominate. Indeed, we find evidence of this: to leading order in $\nu$ and for small $z$ we obtain
\beq
S_\text{out}\simeq -2\pi c (\pi \ell_3 T)^2\,.
\eeq
The dependence $\propto T^2$ is that of a thermal conformal gas in $2+1$ dimensions. In contrast, if we consider solutions with $z\gg 1$ (but with $\nu z\ll 1$), which are close to the global AdS$_3$ vacuum, we get a non-thermal result
\beq
S_\text{out}\simeq -\frac{2\pi}3 c\,.
\eeq

Let us now examine the results for $S_\text{gen}$ and $S_\text{cl}$ more generally, without necessarily restricting to very small $\nu$ nor small $z$.
\begin{figure}
\centering
  \includegraphics[width=.85\textwidth]{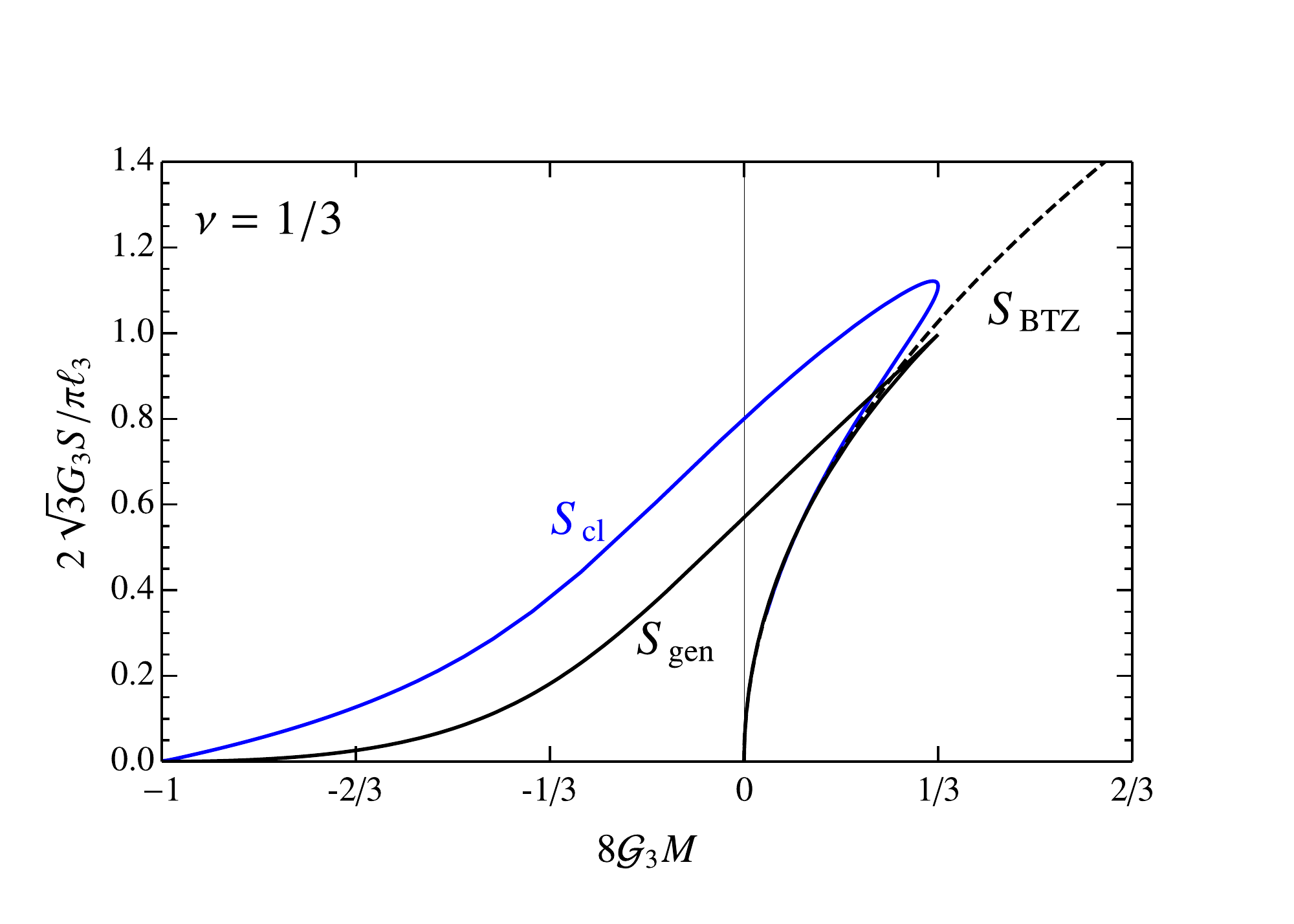}
\caption{\small Quantum and classical entropies of quantum black holes with given mass. The entropy of black holes in branch 3 is $S_\text{BTZ}$. }
\label{fig:S}
\end{figure}
Fig.~\ref{fig:S} shows the entropy of these solutions as a function of the mass $M$ for $\nu=1/3$. As $\nu$ becomes smaller, the entropy curves for $0\leq M \leq 1/(24 \mathcal{G}_3)$ approach that of the BTZ solution, and for $M<0$ they go to zero. Conversely, the differences between curves become larger as $\nu$ grows.

In the range $-1/(8\mathcal{G}_3)< M<0$ we find the quantum-dressed cone solutions of branch $1a$. Since their horizon is due to the backreaction of the Casimir energy in the conical spacetime, their entropy should naturally be interpreted as entanglement entropy of the quantum fields across this horizon \cite{Emparan:2006ni}. In the regime of positive masses, $0\leq M\leq 1/(24 \mathcal{G}_3)$, branch $1b$ solutions have higher entropy than branch 2, which presumably should again be interpreted as mostly due to the entanglement of fields in the state dominated by the Casimir effect. But the duplicity of branches of quBTZ black holes is intriguing. In particular, it would seem that the $M=0$ BTZ solution could develop, through quantum backreaction, a significant non-zero temperature and entropy, dominating entropically over other branches. But, as we saw above, the divergent specific heat of this solution (which is also very large for the solutions with small non-zero $M$) hints at an instability.

More generally, the different dominance in entropy between the branches of quantum black holes and BTZ black strings presumably indicates which phase is preferred (in a microcanonical ensemble). Conventional turning-point arguments indicate that branch 2 solutions should be locally unstable.

One can visually conclude from fig.~\ref{fig:S} that $S_\text{cl}$ cannot satisfy the first law since it has zero and infinite derivatives for certain masses, but we expect that this law works instead for $S_\text{gen}$.
Taking the results for $M$, $T$ and $S_\text{gen}$ in \eqref{Mznu}, \eqref{Tznu} and \eqref{Squst}, and using \eqref{cGnu}, it is straightforward to verify that
\beq
\partial_z M-T \partial_z S_\text{gen}=0
\eeq
for all (fixed) values of $\nu$, so that the first law
\beq\label{stat1law}
dM=TdS_\text{gen}
\eeq
is satisfied by the quantum-corrected entropy, and not by the classical Bekenstein-Hawking-Wald entropy. 

With only the leading curvature correction in the effective theory \eqref{effact}, we should be limited to claiming this result only to cubic order in $\nu$. However, \eqref{stat1law} holds for the quBTZ solutions exactly in $\nu$ when the effects of higher-curvature terms in the mass are resummed with \eqref{exactG}, or its equivalent \eqref{cGnu}. We find this result remarkable, and we will see that it extends to rotating solutions without needing to modify the resummation.

\section{Rotating quBTZ}\label{sec:rotquBTZ}

Now we study the extension of the static AdS C-metric \eqref{statc} to a stationary solution,
\beqa
\label{rotc}
ds^2=\frac{\ell^2}{\lp \ell+x r\rp^2}&\Biggl[& -\frac{H(r)}{\Sigma(x,r)} \lp dt +a x^2d\phi \rp^2+
\frac{\Sigma(x,r)}{H(r)}dr^2\nn\\
&&+r^2\lp \frac{\Sigma(x,r)}{G(x)}dx^2+\frac{G(x)}{\Sigma(x,r)}\lp d\phi-\frac{a}{r^2}dt\rp^2\rp
\Biggr]\,,
\eeqa
where\footnote{In addition to the changes mentioned in footnote~\ref{transl}, relative to \cite{Emparan:1999fd} we have made $\sqrt{\lambda} a\to a/\ell_3$.}
\beqa\label{HGSig}
H(r)&=& \frac{r^2}{\ell_3^2}+\kappa -\frac{\mu\ell}{r}+\frac{a^2}{r^2}\,,\\
G(x)&=& 1-\kappa x^2-\mu x^3+\frac{a^2}{\ell_3^2}x^4\,,\\
\Sigma(x,r)&=& 1+\frac{a^2 x^2}{r^2}\,.
\eeqa
The new parameter $a$, with dimension of length, introduces rotational effects in the spacetime and, without loss of generality, we will take it to be non-negative.

All the other parameters are interpreted as before. The equations for the brane tension, backreaction, and CFT central charge, \eqref{Kab}, \eqref{tension}, \eqref{nu3d}, apply in the same form, and we can place again the brane at $x=0$. The bulk geometry shares many of the features of the Kerr solution, such as a ring singularity at points where $r^2\Sigma=0$, i.e., where
\beq\label{ring}
r=x=0\,.
\eeq

In fact, it is straightforward to show that in the tensionless limit $\ell\to\infty$, making the changes \eqref{toSchAdS} plus $a\to a\ell_3/\ell$, the metric becomes the same as the Kerr-AdS$_4$ solution. On the other hand, in the zero-backreaction limit $\ell\to 0$, where the brane is pushed to the asymptotic boundary of AdS$_4$, the geometry induced at this boundary at $x=0$ is readily seen to have, for $\kappa=-1$, a rotating BTZ black hole in it. In this limit we recover the construction in \cite{Hubeny:2009rc}, which derived the metric as a double Wick rotation of Kerr-AdS$_4$. The details are given in appendix~\ref{app:zerobr}.

The global structure of the rotating solution is more subtle than in the static case. The main features were identified in \cite{Emparan:1999fd}, but the  calculation of the effects on the physical magnitudes was left mostly undone, as was also the dual holographic interpretation.

\subsection{Geometry, $M$, $J$, and $\langle T^a{}_b \rangle$ of quBTZ}\label{subsec:geoqbtz}

We assume that we work in a parameter range where there exists at least one positive root of $G(x)$, the smallest of which we call $x=x_1$ (again, this will be dealt with using an appropriate parametrization). We will then use $x_1$ as the primary parameter instead of $\mu$, which is given by
\beq\label{mux$1a$}
\mu=\frac{1-\kappa x_1^2+\tilde{a}^2}{x_1^3}\,.
\eeq
For convenience we have also introduced another dimensionless parameter for the rotation,
\beq\label{atilde}
\tilde{a}=\frac{a x_1^2}{\ell_3}\,.
\eeq

We will retain (two copies of) the bulk region where $0\leq x\leq x_1$.
The locus $x=x_1$ is a fixed-point set of the Killing vector
\beq\label{rotKV}
\frac{\partial}{\partial\phi}-\tilde{a}\ell_3 \frac{\partial}{\partial t}\,.
\eeq
This has two consequences. First, as before, in order to avoid a conical singularity we must identify $\phi\sim \phi+2\pi\Delta$, with
\beq\label{Deltax$1a$}
\Delta =\frac{2}{|G'(x_1)|}=\frac{2x_1}{3-\kappa x_1^2-\tilde{a}^2}\,.
\eeq
Second, and very importantly, now the identification along orbits of \eqref{rotKV} must be made on spatial surfaces at constant $t+\ell_3\tilde{a}\phi$.
This is a consequence of bulk regularity that would not be visible if we only looked at the `naive brane metric'
\beq\label{naive}
ds^2=-H(r)dt^2+\frac{dr^2}{H(r)}+r^2\lp d\phi-\frac{a}{r^2}dt\rp^2\,,
\eeq
which is induced on the brane at $x=0$ in the coordinates of \eqref{rotc}. Apparently, this would give $M=-\kappa\Delta^2/(8G_3)$ and $J=a\Delta^2/(4G_3)$. However, this is not the globally appropriate form of the rotating quBTZ black hole, since  along an orbit of \eqref{rotKV} $\phi\to\phi+2\pi\Delta$ does not return to the same point in spacetime but to another one at a different $t$. This introduces a rotation of frames that persists even at $r\to\infty$, but we can change to an asymptotically non-rotating frame by shifting $\phi\to \phi + C t$ with an appropriately chosen constant $C$.

Taking all into account, we find that in order to go to canonical coordinates $\bar{t}$ and $\bar{\phi}$ on the asymptotically AdS$_3$ brane we must change
\beqa\label{tphibar}
t&=&\Delta\lp \bar{t}-\tilde{a}\ell_3 \bar{\phi}\rp\,,\nn\\
\phi&=&\Delta \lp \bar{\phi}-\frac{\tilde{a}}{\ell_3}\,\bar{t}\rp\,.
\eeqa
Conversely, the Killing vectors transform as
\beqa\label{canokill}
\frac{\partial}{\partial t}&=&\frac1{\Delta(1-\tilde{a}^2)}\lp \frac{\partial}{\partial\bar{t}}+\frac{\tilde{a}}{\ell_3}\frac{\partial}{\partial\bar{\phi}}\rp\,,\nn\\
\frac{\partial}{\partial \phi}&=&\frac1{\Delta(1-\tilde{a}^2)}\lp \frac{\partial}{\partial\bar{\phi}}+\tilde{a}\ell_3\frac{\partial}{\partial\bar{t}}\rp\,.
\eeqa
In addition, we also redefine
\beq\label{rtorbar}
r^2=\frac{\bar{r}^2-r_s^2}{(1-\tilde{a}^2) \Delta^2}
\eeq
with
\beq\label{ringbarr}
r_s=\ell_3\frac{\tilde{a}\Delta}{x_1}\sqrt{2-\kappa x_1^2}=\ell_3\frac{2\tilde{a}\sqrt{2-\kappa x_1^2}}{3-\kappa x_1^2 -\tilde{a}^2}\,.
\eeq

The metric of the rotating quBTZ black hole now has points identified along orbits of $\partial/\partial\bar{\phi}$, with $(\bar{t},\bar{\phi})\sim (\bar{t},\bar{\phi}+2\pi)$, and the geometry is asymptotically AdS$_3$,
\beqa\label{rotquBTZ}
ds^2&=& -\lp \frac{\bar{r}^2}{\ell_3^2}-8\mathcal{G}_3 M -\frac{\ell\mu\Delta^2}{r}\rp d\bar{t}^2+\lp \bar{r}^2+\ell_3^2\frac{\ell\mu \tilde{a}^2\Delta^2 }{r}\rp d\bar{\phi}^2\nn\\
&& -8 \mathcal{G}_3 J\lp 1 +\frac{\ell}{x_1 r}\rp d\bar{t}d\bar{\phi}
\nn\\
&&+\lp\frac{\bar{r}^2}{\ell_3^2}-8\mathcal{G}_3 M+\frac{(4\mathcal{G}_3 J)^2}{\bar{r}^2}-\ell\mu (1-\tilde{a}^2)^{2}\Delta^4\frac{r}{\bar{r}^2}\rp^{-1}d\bar{r}^2\,,
\eeqa
where $r$ is now a function of $\bar{r}$, shorthand for \eqref{rtorbar}. The `renormalized' three-dimensional Newton's constant $\mathcal{G}_3$ was defined in \eqref{exactG}. We have
\begin{align}\label{MquBTZ}
M&=- \frac{\kappa}{8\mathcal{G}_3}\Delta^2\lp 1+\tilde{a}^2-\frac{4\tilde{a}^2}{\kappa x_1^2}\rp\nn\\
&=\frac1{2\mathcal{G}_3}\frac{-\kappa x_1^2 +\tilde{a}^2(4-\kappa x_1^2)}{\lp 3-\kappa x_1^2 -\tilde{a}^2\rp^2}
\end{align}
and
\begin{align}\label{JquBTZ}
J&=\frac{\ell_3}{4\mathcal{G}_3}\tilde{a}\mu x_1\Delta^2\nn\\
&=\frac{\ell_3}{\mathcal{G}_3} \frac{\tilde{a}(1-\kappa x_1^2+\tilde{a}^2)}{\lp 3-\kappa x_1^2 -\tilde{a}^2\rp^2}
\end{align}
(note that these depend on $\tilde a$ and $\kappa x_1^2$, and not on $\kappa$, $x_1$, or $a/\ell_3$ separately).

In this manner it is apparent that in the limit $\ell\to 0$, in which the quantum backreaction vanishes, the metric \eqref{rotquBTZ} for $\kappa=-1$ is the same as the classical rotating BTZ solution with mass $M$ and angular momentum $J$ (and for $\kappa=+1$ we obtain rotating conical defects). The terms $\propto \ell$ decay faster at $\bar{r}\to\infty$ than the asymptotic terms from which the mass and angular momentum are read, so these are indeed given by \eqref{MquBTZ} and \eqref{JquBTZ} for all $\ell$. For BTZ black holes it is useful to know that
\beq\label{MpmJ}
8\mathcal{G}_3\lp M\pm \frac{J}{\ell_3}\rp=\frac{4(1-\tilde{a}^2)(-\kappa x_1^2\pm 2\tilde{a}^2)}{\lp 3-\kappa x_1^2 -\tilde{a}^2\rp^2}
\,.
\eeq

The $\ell=0$ limit also clarifies the transformations \eqref{tphibar} and \eqref{rtorbar} made above, since these transformations act on a BTZ black hole to yield another BTZ black hole with different $M$ and $J$. The specific choice \eqref{tphibar} is selected by the identification of points imposed by bulk regularity. It turns out that the same transformations work to bring the metric into the correct form independently of $\ell$. That is, the geometry involves the same global aspects in the $(x,\phi)$ sector regardless of the strength of the backreaction.

Since the curvature singularity \eqref{ring} lies entirely on the brane, the rotating quBTZ metric \eqref{rotquBTZ} possesses a ring singularity at $r=0$, that is, at $\bar{r}=r_s$, \eqref{ringbarr},
which was not present in the classical rotating BTZ geometry.

Like in the Kerr and rotating BTZ solutions, there exist regions of the spacetime with closed timelike curves, and avoiding naked ones requires parameter restrictions. For instance, in the following we will impose $\tilde{a}\leq 1$ so that in \eqref{canokill} and \eqref{rtorbar} we have $1-\tilde{a}^2\geq 0$.
Other parameter constraints come from requiring that $\Delta>0$. This is  satisfied as long as $-\kappa x_1^2>\tilde{a}^2-3$, which henceforth we will also assume (for $\kappa=-1$ and $\tilde{a}^2\leq 1$ it always holds). However, the detailed study of these and related constraints is beyond the scope of this article. The black hole horizons and their properties will be examined later.

In order to obtain the holographic stress tensor, it is simpler to start with its form  (as read from the 3D gravitational equation \eqref{3Deffeqn}) in the coordinates $(t,r,\phi)$ of the `naive metric' \eqref{naive}. We only give the results for the leading order piece $\langle T^{a}{}_{b} \rangle_0$, which we obtain from \eqref{Tab0}, yielding
\beqa
\langle T^{t}{}_{t} \rangle_0&=&\langle T^{r}{}_{r} \rangle_0=-\frac12\langle T^{\phi}{}_{\phi} \rangle_0=\frac1{16\pi G_3}\frac{\ell\mu}{r^3}\,,\nn\\
\langle T^{\phi}{}_{t} \rangle_0&=&\frac1{16\pi G_3}\frac{3\ell\mu a}{r^5}\,,\label{naiveT}
\eeqa
 and then change to $(\bar{t},\bar{r},\bar{\phi})$ to find
\beqa
8\pi G_3\langle T^{\bar{t}}{}_{\bar{t}} \rangle_0&=&\frac{\ell\mu}{2(1-\tilde{a}^2)\, r^3}\lp 1+2\tilde{a}^2+\frac{3\tilde{a}^2\ell_3^2}{x_1^2 r^2}\rp\,,\label{Ttt}\\
8\pi G_3\langle T^{\bar{\phi}}{}_{\bar{\phi}} \rangle_0&=&-\frac{\ell\mu}{2(1-\tilde{a}^2)\, r^3}\lp 2+\tilde{a}^2+\frac{3\tilde{a}^2\ell_3^2}{x_1^2 r^2}\rp\,,\\
8\pi G_3\langle T^{\bar{t}}{}_{\bar{\phi}} \rangle_0&=&-\ell_3\frac{3\ell\mu \tilde{a}}{2(1-\tilde{a}^2)\, r^3}\lp1+\frac{\tilde{a}^2\ell_3^2}{x_1^2 r^2}\rp\,,\\
8\pi G_3\langle T^{\bar{\phi}}{}_{\bar{t}} \rangle_0&=&\frac1{\ell_3}\frac{3\ell\mu \tilde{a}}{2(1-\tilde{a}^2)\, r^3}\lp1+\frac{\ell_3^2}{x_1^2 r^2}\rp\,,\\
8\pi G_3\langle T^{\bar{r}}{}_{\bar{r}} \rangle_0&=&\frac{\ell\mu}{2r^3}\label{Trr}\,.
\eeqa
Here again $r$ stands for \eqref{rtorbar}.
Using \eqref{nu3d} or \eqref{smallnu3d} we can express the metric and the stress tensor in terms of only 3D magnitudes, namely, $c$, $\ell_3$, $G_3$, $M$ and $J$.\footnote{In appendix \ref{app:TabBTZ} we give the stress tensor in terms of other parameters more closely related to the BTZ geometry.}

Recall now that we are using $\ell$, $\ell_3$, $\kappa x_1^2$ and $\tilde{a}$ as parameters, in terms of which all other quantities are obtained using \eqref{mux$1a$} and \eqref{Deltax$1a$}. We regard $\ell_3$ as fixing the scale of the geometry, while $\ell/\ell_3$ measures the strength of the backreaction through \eqref{nu3d}. We see that $M$ and $J$ in \eqref{MquBTZ} and \eqref{JquBTZ} depend on $\kappa x_1^2$ and $\tilde{a}$ but not on $\ell$. Moreover, $\ell$ enters in \eqref{Ttt}-\eqref{Trr} only as an overall prefactor. This implies, again, that the dependence of $\langle T^{a}{}_{b} \rangle_0$ on $\mathcal{G}_3M$ and $\mathcal{G}_3J$ is unaffected by the strength of the backreaction; in other words, $\langle T^{a}{}_{b} \rangle_0$ depends on backreaction only through $\mathcal{G}_3$ \eqref{approxG}.

For this solution the dependence of the stress tensor on $M$ and $J$ cannot be characterized by a single function as in the static case, but for our purposes it will suffice to consider the asymptotic leading term at large $\bar{r}$ in the energy density. After replacing \eqref{rtorbar} in \eqref{Ttt}, we define
\begin{align}\label{FMJ}
F(M,J)&=\frac{\mu \Delta^{3}\sqrt{1-\tilde{a}^2}}{2}\lp 1+2\tilde{a}^2\rp\nn\\
&=8\frac{\sqrt{1-\tilde{a}^2}(1+2\tilde{a}^2)(1-\kappa x_1^2+\tilde{a}^2)}
{\lp 3-\kappa x_1^2 -\tilde{a}^2\rp^3}
\,.
\end{align}
Essentially the same function, up to simple factors, controls the large $\bar{r}$ asymptotics of all other components of the stress tensor. It vanishes for $\kappa x_1^2=1+\tilde{a}^2$, which is equivalent to $\mu=0$ and corresponds to the empty global AdS$_3$ solution. It also vanishes for $\tilde{a}^2=1$, which, as \eqref{MpmJ} shows, are solutions with $M\pm J/\ell_3=0$. We will examine these next.

\subsection{Branches of solutions and bounds on $M$ and $J$}

Like in the static case, for certain values of $(M,J)$ there exist different branches of solutions. We are mostly interested in the regime of non-negative masses, but much of the analysis can be made without this restriction. We consider $J\geq 0$ without loss of generality.

\begin{figure}[t]
\centering
	\includegraphics[width=.56\textwidth]{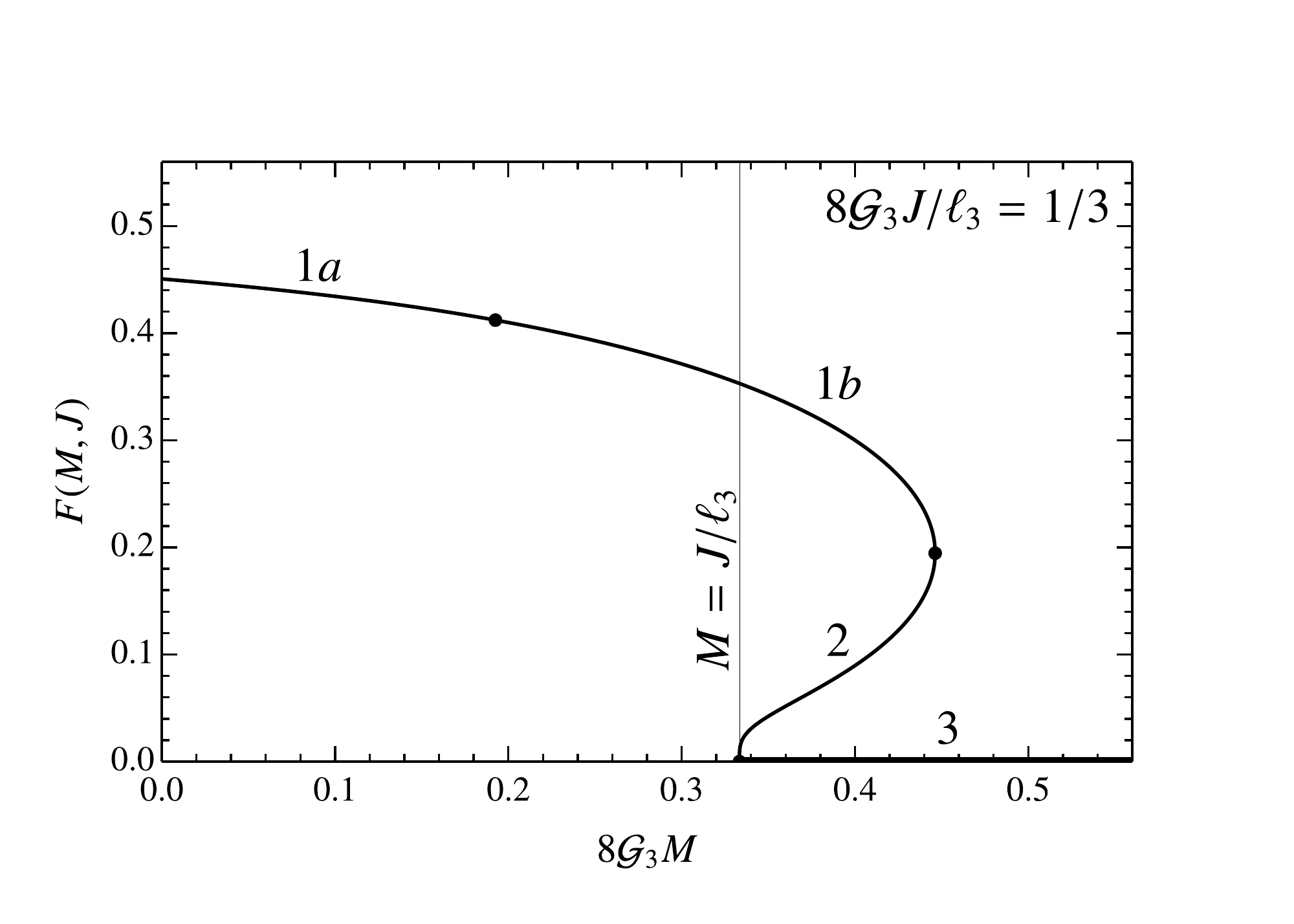}
	\includegraphics[width=.43\textwidth]{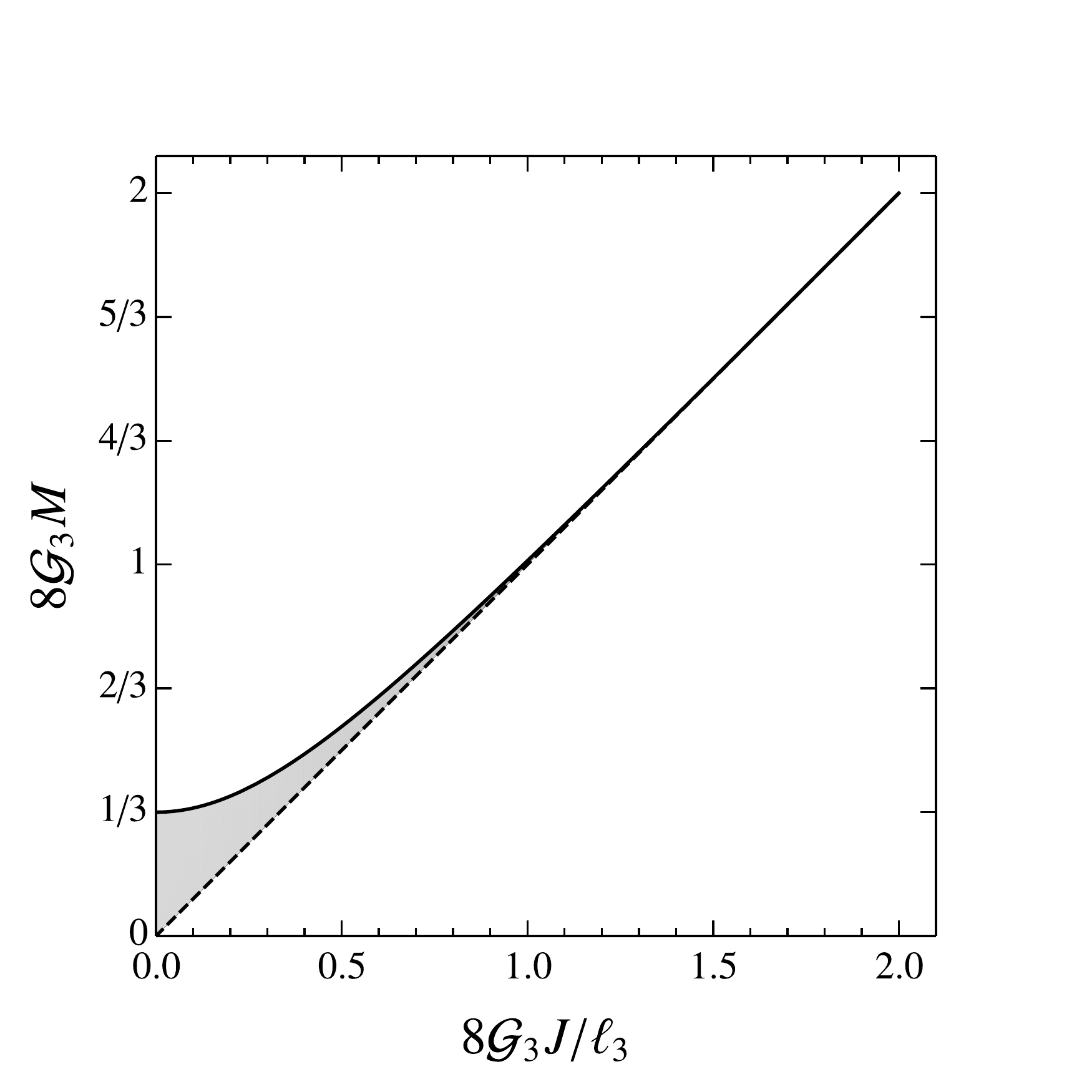}
\caption{\small Left: the asymptotic energy density function $F(M,J)$ \eqref{FMJ} for a fixed value of $J$ representative of $0<J<\ell_3/(4 \mathcal{G}_3)$ (we only show $M\geq 0$ but branch $1a$ solutions extend to $M<0$). Branches $1a$ and $1b$ have large quantum effects that enable them to exist with $M\leq J/\ell_3$.  Branch 3 are rotating BTZ black strings. For $J\geq \ell_3/(4 \mathcal{G}_3)$ branch 2 disappears and all the solutions have masses below the classical BTZ extremal limit $M=J/\ell_3$. Right: in gray, the range of masses and angular momenta for which the quBTZ solutions exist above the classical extremal bound (lower dashed line). The upper solid curve is the bound on the masses implied by the holographic construction \eqref{maxMJ}. For each point in the gray region there are solutions in branches $1b$ and 2. Branch 1 solutions also exist below the classical extremal line, but whether they are free of pathologies or not depends on the strength of the backreaction, and is not fully investigated in this article.}
\label{fig:range}
\end{figure}

A plot of $F(M,J)$ at constant $J>0$ (not larger than a value that we discuss below), see fig.~\ref{fig:range} (left), reveals the existence of branches of solutions analogous to the ones we found in the static case. Branches $1a$ and $1b$ are distinguished by the sign of $\kappa$ but otherwise smoothly continue into each other.
We observe a maximum value of $M$ for fixed $J$, where branches $1b$ and 2 meet, and a minimum $M$ along branch 2.

We can easily determine these special values by extremizing $M$ for fixed $J$. We find two classes of solutions, both of them with $\kappa=-1$. The first are minima and correspond to
\beq\label{extBTZ}
\tilde{a}=1\,,\qquad M=\frac{J}{\ell_3}=\frac{1}{\mathcal{G}_3(2+x_1^2)}\,.
\eeq
This is the classical extremality bound for BTZ.
As we noticed above, the stress-energy tensor vanishes identically in these solutions. Thus they must be regarded as the rotating extensions of the static $M=0$ solution where branch 2 reaches its minimum mass.

The second class of extrema are maxima and occur for
\beq\label{maxMJ}
x_1^2+\tilde{a}^2=3\,,\qquad M=\frac1{8\mathcal{G}_3}\lp \frac{12}{x_1^4}-1\rp\,,\qquad J=\frac{\ell_3}{\mathcal{G}_3}\frac{\sqrt{3-x_1^2}}{x_1^4}\,.
\eeq
When $\tilde{a}=0$ and $x_1=\sqrt{3}$, this reproduces the static upper bound $M= 1/(24 \mathcal{G}_3)$. These are the solutions where branches $1b$ and 2 meet at the maximum $M$ for fixed $J$. Adding rotation increases the maximum mass, but for $x_1=\sqrt{2}$ the extremal bound \eqref{extBTZ} is reached. This corresponds to
\beq\label{upperMJ}
M=\frac{J}{\ell_3}=\frac1{4\mathcal{G}_3}\,.
\eeq
For $M$ and $J$ above these values, solutions still exist but branch 2 disappears. All these branch 1 solutions have $M\leq J/\ell_3$, and the branch ends at the extremal solution with $F=0$.

The region between \eqref{extBTZ} and \eqref{maxMJ}, with branch $1b$ and branch 2 solutions that satisfy the classical extremal bound $M\geq J/\ell_3$, is depicted in fig.~\ref{fig:range} (right).

For any value of $J$, whenever $ -\kappa x_1^2<2\tilde{a}^2$ the classical extremal bound is violated by branch 1 black holes (see \eqref{MpmJ}).
Whether these black holes are non-pathological---without naked singularities nor naked closed timelike curves---relies on the existence and location of horizons. Unlike the study so far, these depend on the backreaction parameter $\ell$. Although we will make a few more observations in the next subsection, a complete study of the regions of the plane $(J,M)$ where physically sensible black holes exist is beyond the scope of this article.
Nevertheless, we anticipate that, since the static black holes with $-1/(8 \mathcal{G}_3)<M<0$ in branch $1a$ are free from any pathologies, it is natural to expect that in some ranges the rotating black holes with $M< J/\ell_3$ will also be valid, made possible by the presence of significant quantum backreaction effects. One of these effects, as we will see, is to make the branch $1b$ solution with $M=J/\ell_3$ have non-zero entropy and temperature.

Rotating BTZ black strings (branch 3) exist in the bulk for all values of $M$ and $J$ satisfying $M\geq J/\ell_3$.  These solutions are natural candidates for the holographic description of the (unexcited) CFT on the rotating BTZ backgrounds in the cases where the quBTZ black hole does not exist or is subdominant. Like all black strings in Karch-Randall braneworlds, they require an infrared regulator.

\subsection{Comparison to other calculations}

The renormalized stress tensor for a free CFT in the classical rotating BTZ geometry takes the form \cite{Steif:1993zv,Casals:2019jfo}
\beqa
8\pi G_3\langle T^{\bar{t}}{}_{\bar{t}} \rangle&=&\sum_{n=1}^\infty \frac{1}{r_n^3}\lp A_n+\frac{\tilde{A}_n}{r_n^2}\rp
\,,\label{fTtt}\\
8\pi G_3\langle T^{\bar{\phi}}{}_{\bar{\phi}} \rangle&=&-\sum_{n=1}^\infty \frac{1}{r_n^3}\lp B_n+\frac{\tilde{A}_n}{r_n^2}\rp\,,\label{fTphiphi}\\
8\pi G_3\langle T^{\bar{t}}{}_{\bar{\phi}} \rangle&=&-\ell_3\sum_{n=1}^\infty \frac{1}{r_n^3}\lp E_n+\frac{\tilde{E}_n}{r_n^2}\rp\,,\label{fTtphi}\\
8\pi G_3\langle T^{\bar{\phi}}{}_{\bar{t}} \rangle&= &\frac1{\ell_3}\sum_{n=1}^\infty \frac{1}{r_n^3}\lp E_n+\frac{\tilde{F}_n}{r_n^2}\rp
\,,\label{fTphit}\\
8\pi G_3\langle T^{\bar{r}}{}_{\bar{r}} \rangle&=&\sum_{n=1}^\infty \frac{C_n}{r_n^3}\,,\label{fTrr}
\eeqa
where
\beq
r_n=\sqrt{D_n\bar{r}^2+\tilde{D}_n}\,,
\eeq
and the coefficients $A_n$,  $\tilde{A}_n$, $B_n$, \dots (with $A_n-B_n+C_n=0$ for tracelessness) are fairly complicated functions of $M$ and $J$ which can be found in \cite{Casals:2019jfo}. The sums are again a consequence of the construction with the method of images.

The first thing to observe  is that, except for the specific form of the coefficients, each individual summand has a dependence on $\bar{r}$ with precisely the same structure as in the holographic result \eqref{Ttt}-\eqref{Trr}. This coincidence is remarkable since it is not directly implied by conformal symmetry. One curious feature is that both results can be formally obtained by an $SL(2,\R)$ transformation from a simple `naive' geometry where $\langle T^{t}{}_{\phi}\rangle=0$. For the free field, this is the static BTZ solution with $8G_3 M=1$ \cite{Steif:1993zv}.

Nevertheless, and crucially, the dependence on $\bar{r}$ of the \textit{total sum} is much more complicated than that of individual summands. In the static case, the sum over images only entered in the overall coefficient $F(M)$, while the radial dependence remained $1/\bar{r}^3$ in both approaches. However, when rotation is present, the holographic CFT has a much simpler radial dependence than the free conformal field. In this instance it does not make much sense to compare the individual coefficients $A_n$ etc.\ to the coefficients in the holographic result (although some aspects of the dependence on $M$ and $J$ can be compared, see appendix~\ref{app:TabBTZ}). This difference in the stress tensors also implies that the backreaction corrections to the metric for the free field will depend on $\bar{r}$ in a much more complicated manner than in the holographic calculation.

One consequence of this more complex radial dependence is that, while the holographic stress tensor is manifestly non-singular everywhere outside the ring singularity at \eqref{ringbarr}, and in particular at the inner Cauchy horizon of BTZ, in the case of the free field this is a much more delicate matter. While the infinite sum leads to a divergent pile up behind the Cauchy horizon \cite{Casals:2019jfo}, there is no divergence when it is approached from the outside \cite{Dias:2019ery,Hollands:2019whz}. We will return to this point in the final discussion.

The holographic stress tensor of the CFT in a non-dynamical BTZ background can be obtained with the construction in \cite{Hubeny:2009rc}. As shown in appendix~\ref{app:zerobr}, the bulk metric employed there is recovered as the limit $\ell\to 0$ of our construction. Since we have found that the stress-energy tensor \eqref{Ttt}-\eqref{Trr} depends on $\ell$ only through an overall prefactor $\ell/G_3\propto c$, these expressions also yield the correct stress-energy tensor for the holographic construction in \cite{Hubeny:2009rc}.

\subsection{Quantum black hole thermodynamics}

We now turn to the analysis of the black hole horizon and its thermodynamics. We will assume that we are in a parameter range where there exists a positive root $r_+$ of $H(r)$, which is a horizon of the Killing vector
\beq\label{horgen}
k=\frac{\partial}{\partial t}+\frac{a}{r_+^2}\frac{\partial}{\partial\phi}\,.
\eeq
It is easy to see that when $a\neq 0$ this horizon is in general accompanied by an inner horizon at $r=r_-<r_+$, as is expected of rotating black holes. For the most part we will only consider the outer event horizon, even though much of the analysis formally applies to the inner one too.

Once again, we resort to a different parametrization. In addition to $z$ and $\nu$ in \eqref{zdef} and \eqref{nudef}, we introduce a new parameter $\alpha$ for the rotation, defined as
\beq\label{alphadef}
\alpha=\frac{a x_1}{\sqrt{-\kappa}\,\ell_3}=\frac{\tilde{a}}{\sqrt{-\kappa}\,x_1}\,.
\eeq
The factor $\sqrt{-\kappa}$ is included since it eliminates $\kappa$ from all physical magnitudes below. Since we are primarily interested in the case $\kappa=-1$, this is inconsequential. For $\kappa=+1$ it should imply that $\alpha^2<0$. These are the branch $1a$ negative mass quantum-dressed cones, where rotation can give rise to naked CTCs. Although it should be interesting to explore this sector of the solutions more carefully, we will not pursue it here.

We can now express all physical magnitudes in terms of the dimensionless parameters $\nu$, $z$, $\alpha$, plus the scales $\ell_3$ and $G_3$ (or $\mathcal{G}_3$), using that
\beqa
x_1^2&=&-\frac1{\kappa}\frac{1-\nu z^3}{z^2\lp 1+\nu z-\alpha^2 z(z-\nu)\rp}\,,\\
r_+^2&=&-\ell_3^2\kappa \frac{1+\nu z-\alpha^2 z(z-\nu)}{1-\nu z^3}\,,\\
\mu x_1 &=& -\kappa\frac{(1+z^2)\lp 1+\alpha^2(1-z^2)\rp}{1-\nu z^3}\,.
\eeqa
Our calculations can be formally carried out for any values of the parameters, but they only apply to black holes as long as $r_+^2$ is positive and real. If we require, as we did in the absence of rotation, that $\text{sign}(\nu z^3-1)=\kappa$, then we see that the rotation parameter is bounded above,
\beq\label{alphabound}
\alpha^2\leq \frac{1+\nu z}{z(z-\nu)}\,,
\eeq
which can be thought of as the analogue of the Kerr bound. In the case $\kappa=-1$ with $\alpha^2>0$ and $\nu<z<\nu^{-1/3}$, this bound also implies that $1+\alpha^2(1-z^2)>0$, so $\mu>0$. On the other hand, the classical extremal limit \eqref{extBTZ} corresponds to $\alpha^2=1/(-\kappa x_1^2)$, that is,
\beq\label{alphaext}
\alpha^2=\frac{z^2(1+\nu z)}{1-2\nu z^3+z^4}\,.
\eeq
Recall, however, that these solutions lie in branch 2, but there are branch $1b$ solutions with the same values of $M=J/\ell_3$.

After some algebra we find that \eqref{MquBTZ} and \eqref{JquBTZ} now take the form
\beq
M=\frac1{2\mathcal{G}_3}\frac{(1-\nu z^3)\lp z^2(1+\nu z)+\alpha^2\lp 1+4z^2+4(1+\alpha^2)\nu z^3-(1+4\alpha^2)z^4\rp\rp}{\lp 1+3z^2+2\nu z^3-\alpha^2(1-4\nu z^3+3z^4)\rp^2}\,,
\eeq
and
\beq
J=\frac{\ell_3}{\mathcal{G}_3}\frac{\alpha z(1+z^2)(1+\alpha^2(1-z^2))\sqrt{(1-\nu z^3)(1+\nu z-\alpha^2 z(z-\nu))}}{\lp 1+3z^2+2\nu z^3-\alpha^2(1-4\nu z^3+3z^4)\rp^2}\,.
\eeq

The horizon is generated by the orbits of the Killing vector \eqref{horgen}, but, using \eqref{canokill}, the canonically normalized generator is instead
\beq
\bar{k}=\frac{\Delta(1-\tilde{a}^2)}{1+\frac{a^2 x_1^2}{r_+^2}}\,k=
\frac{\partial}{\partial \bar{t}}+\Omega\frac{\partial}{\partial\bar{\phi}}
\eeq
where the horizon angular velocity is
\beqa
\Omega&=&\frac{a}{\ell_3^2}\frac{\ell_3^2+r_+^2 x_1^2}{r_+^2+a^2 x_1^2}\nn\\
&=&\frac1{\ell_3}\frac{\alpha (1+z^2)\sqrt{(1-\nu z^3)(1+\nu z-\alpha^2 z(z-\nu))}}{z(1+\nu z)\lp 1+\alpha^2(1-z^2)\rp}\,.
\eeqa
Relative to $\bar{k}$, the horizon temperature is
\begin{align}
T&=\frac{\Delta(1-\tilde{a}^2)}{1+\frac{a^2 x_1^2}{r_+^2}}\frac{H'(r_+)}{4\pi}\nn\\
&=\frac1{2\pi\ell_3}
\frac{\lp z^2(1+\nu z)-\alpha^2(1-2\nu z^3+z^4)\rp
\lp 2+3(1+\alpha^2)\nu z-4\alpha^2 z^2+\nu z^3+\alpha^2 \nu z^5\rp}{z(1+\nu z)\lp 1+\alpha^2(1-z^2)\rp\lp 1+3z^2+2\nu z^3-\alpha^2(1-4\nu z^3+3z^4)\rp}\,.
\end{align}

The conditions under which this $T$ is non-negative are complicated. The temperature vanishes for extremal solutions with $\alpha$ as in \eqref{alphaext}, due to the vanishing factor $1-\tilde{a}^2$ (instead of because $H'=0$). This was expected, since the stress tensor in these solutions vanishes and does not backreact. In contrast, the temperature is non-zero for branch $1b$ solutions with $M=J/\ell_3$ (just like it is for the branch $1b$ static $M=0$ black hole), and continues to be positive along that branch for a range of $M<J/\ell_3$. By itself, the bound \eqref{alphabound} does not seem to guarantee that $T$ is not negative, even for $\kappa=-1$ and $\nu<z<\nu^{-1/3}$. Further parameter restrictions may be necessary, but we will not undertake their analysis here.

Finally, the area of the bulk horizon at $r=r_+$, in units of $4G_4$, yields the holographic quantum entropy. Taking into account  the change in \eqref{tphibar}, we find
\beqa\label{SquBTZr}
S_\text{gen}&=&\frac1{2G_4}\int_0^{2\pi}d\bar{\phi}\int_0^{x_1}dx\frac{r_+^2\ell^2}{\lp \ell+r_+ x\rp^2}\Delta\lp 1+\frac{a^2 x_1^2}{r_+^2}\rp\nn\\
&=& \frac{\pi}{G_4}\Delta\frac{\ell x_1\lp r_+^2+a^2 x_1^2\rp}{\ell+r_+ x_1}\nn\\
&=& \frac{\pi\ell_3}{G_3}\frac{z\lp 1+\alpha^2(1-z^2)\rp\sqrt{1+\nu^2}}{1+3z^2+2\nu z^3-\alpha^2(1-4\nu z^3+3z^4)}\,.
\eeqa

These thermodynamic expressions are fairly complicated, but one can verify by explicit calculation that,  if the higher-curvature corrections to $M$ and $J$ are exactly resummed using \eqref{cGnu}, then
\beqa\label{dzdaM}
\partial_z M-T\partial_z S_\text{gen}-\Omega \partial_z J&=&0\,,\nn\\
\partial_\alpha M-T\partial_\alpha S_\text{gen}-\Omega \partial_\alpha J&=&0\,,
\eeqa
which amounts to proving that the quantum entropy $S_\text{gen}$ correctly satisfies the first law
\beq\label{1stlaw2}
d M-Td S_\text{gen}-\Omega d J=0
\eeq
for all $\nu$, that is, for all values of the brane tension and strength of backreaction. Indeed, it is formally satisfied for all values of the parameters, regardless of whether the solutions are physically sensible or not.

For its part, the Bekenstein-Hawking entropy of the horizon on the brane is
\beqa
S_\text{cl}&=&\frac{2\pi r_+}{4 G_3}\Delta\lp 1+\frac{a^2 x_1^2}{r_+^2}\rp\nn\\
&=& \frac{1+\nu z}{\sqrt{1+\nu^2}}S_\text{gen}\,.
\eeqa
In the limit $\nu\to 0$ in which the backreaction disappears, we correctly recover the classical BTZ result,
\beq
S_\text{gen}|_{\nu=0}= S_\text{cl}|_{\nu=0}= S_\text{BTZ}=
\frac{\pi\ell_3}{\sqrt{2G_3}}\lp \sqrt{M+\frac{J}{\ell_3}} + \sqrt{M-\frac{J}{\ell_3}}\rp \,.
\eeq
The Wald entropy corrections are straightforward to compute but they are not particularly illuminating so we omit them. The part of the quantum entropy due to entanglement of CFT fields again satisfies \eqref{Sout} to leading order in the backreaction.

\section{Discussion and outlook}\label{sec:discuss}

Our study of holographic quBTZ has refined and extended the early analysis in \cite{Emparan:2002px} in several important ways. Our main results are the metric of quBTZ \eqref{rotquBTZ}, its renormalized CFT stress tensor \eqref{Ttt}-\eqref{Trr}, and its quantum entropy \eqref{SquBTZr} which satisfies the first law \eqref{q1stlaw}.
For our goal of obtaining these and other physical magnitudes of the solutions we have not needed to know the parameter ranges for which the rotating geometries are free of pathologies, but it should be interesting to understand them better.

We have also properly identified the effects of backreaction, and accounted for the leading higher-curvature corrections to the three-dimensional effective theory \eqref{effact}. These terms are the same as in the new massive gravity of \cite{Bergshoeff:2009hq}, a feature which to our knowledge has not been noticed before, and which can likely be connected to the perturbative analysis of braneworld massive gravity \cite{Karch:2000ct}.

Another aspect that deserves further investigation is the dynamical and thermodynamical stability of quBTZ black holes. We have found intriguing features in the duplicity of branches of quantum black holes with the same mass but with very different temperature, entropy, and specific heat. The latter, and in particular its sign, may indicate how the black hole exchanges radiation with the CFT in the non-dynamical part of the AdS$_4$ boundary.

One important conclusion is that (barring the existence of other bulk solutions for localized black holes) holographic quantum effects in BTZ are important only up to a maximum mass, not larger than $1/(24 G_3)$ in the static case and than $J/\ell_3=1/(4G_3)$ with rotation. In this range of masses the quantum effects are captured in an exact, analytic manner by the AdS C-metric braneworld solutions. For larger masses, the only known phases correspond to bulk BTZ black strings (with an infrared regulator), which imply the complete suppression of quantum effects on BTZ black holes with these masses. This may seem surprising, but it is not uncommon that the holographic description of CFTs in black hole backgrounds gives results at odds with the expectations from weakly coupled fields \cite{Marolf:2013ioa}. As we have argued, this upper mass limit is likely a feature also of holographic quantum effects in AdS black holes in higher dimensions.

To finish, we discuss a few other remaining issues and possibilities for further investigation.

\paragraph{Quantum bulk backreaction and strong cosmic censorship.} In our study the bulk is treated classically, which corresponds to the leading order limit of the CFT in an expansion for large central charge $c$. Quantum bulk physics then gives corrections in inverse powers of $c$. Our classical bulk is qualitatively very similar to the Kerr-AdS$_4$ black hole, and so we expect that perturbative bulk  quantum corrections will be qualitatively like in Kerr-AdS$_4$. That is, the bulk black hole geometry will receive small corrections almost everywhere, in particular at the outer event horizon, but the effects of the bulk quantum stress tensor will become large, indeed divergent, when the inner Cauchy horizon is approached from the outside \cite{Hollands:2019whz}. This reasoning has been employed in \cite{Emparan:2020rnp} in order to argue that strong cosmic censorship is upheld in the BTZ black hole. The question has acquired interest recently, after it has been shown in \cite{Dias:2019ery,Papadodimas:2019msp,Balasubramanian:2019qwk,Hollands:2019whz} that leading order perturbative quantum effects do not spoil the regularity of the Cauchy horizon of BTZ. However, \cite{Emparan:2020rnp} argues that this smoothness will not survive effects at the next perturbative order. In our holographic construction, the inner horizon of rotating quBTZ is smooth, but bulk quantum effects will act on it as they do for Kerr black holes. That is, they will enforce strong cosmic censorship in the bulk black hole, and consequently, also in the quBTZ black hole on the brane.\footnote{It seems unlikely to us that the presence of the brane can alter this conclusion. In the most extreme limit of a tensionless brane, it simply acts as a $\Z_2$ projection on bulk fields. The divergence of the stress tensor  present in that case should not disappear when the bulk solution becomes less symmetric with tensional branes. We acknowledge discussions with Jorge Santos on this point.}  Let us emphasize that the presentation of this argument in \cite{Emparan:2020rnp} is unaffected by the fact that it did not account for the global effects on rotating quBTZ discussed in sec.~\ref{subsec:geoqbtz}.

\paragraph{Classical holographic proofs for quantum entropies.} We have proven through a direct calculation that the first law of quantum black holes holds for the specific quBTZ solutions, but it seems very likely that a general holographic proof of \eqref{q1stlaw} should be possible using only classical theorems in the bulk, without any explicit solutions. Such a derivation should indeed apply in generic braneworld holography, including asymptotically flat Randall-Sundrum branes, where it has also been verified in explicit solutions \cite{Emparan:1999wa}. In Karch-Randall constructions with AdS branes, it should illuminate the simple exact resummation that we have found for the higher-curvature corrections to the mass and angular momentum.

As we have mentioned, this holographic quantum first law is not manifestly the same as a `bulk first law' since $M$ and $J$ must be defined with reference to the brane geometry and to the effective three-dimensional theory. In flat-bulk (ADD) braneworlds, the bulk mass and spin agree with their brane counterparts \cite{Emparan:2000rs}, and the same result should apply in Randall-Sundrum asymptotically flat braneworlds, owing to the fact that they recover the lower dimensional gravitational field of a point mass, without any modifications from higher-curvature corrections. Similar reasoning may work in Karch-Randall braneworlds, with due care of higher-curvature effects on mass and spin. Then one should prove that these $M$ and $J$, with the non-standard asymptotics of the bulk, satisfy a first law for the bulk entropy. This seems doable.\footnote{The analysis of the thermodynamics of the AdS C-metric in \cite{Anabalon:2018ydc}, although potentially related, does not immediately apply since it does not include the brane, and it is not obvious that the bulk mass and spin defined there are the same as the mass and spin defined on the brane that enter in \eqref{q1stlaw}. }

More broadly, other theorems for quantum entropies of solutions of the equations \eqref{semieins} can be readily proven using holography. For instance, the second law for $S_\text{gen}$, proven in \cite{Wall:2011hj} in the perturbative expansion in $\hbar$, is in the holographic set up an immediate consequence of the bulk second law; in fact, it is not a perturbative but an exact statement within the planar limit of the CFT. It is also interesting to consider the bulk view of how $S_\text{cl}$ will not in general satisfy a second law. For instance, if a black hole localized on the brane begins to slip away from the brane, the area of the brane horizon, and hence $S_\text{cl}$, will decrease, while the classical bulk evolution guarantees that $S_\text{gen}$ will grow.\footnote{Then the radiation entropy $S_\text{out}$ will grow. This is intriguingly suggestive of a classical bulk dual of Hawking evaporation \cite{Tanaka:2002rb,Emparan:2002px}, but so far this picture has resisted attempts at being further substantiated (see \cite{Figueras:2011gd}). RE thanks Nemanja Kaloper and Takahiro Tanaka for very many conversations on this topic.}

Along these same lines, one can also envisage holographic proofs of theorems for quantum extremal surfaces by adapting theorems of classical extremal surfaces. For instance, the classical theorem that the apparent horizon lies inside the event horizon will, under suitable conditions, imply that quantum extremal surfaces on the brane are covered by event horizons.

\paragraph{Extended thermodynamics of quantum BTZ.} In \eqref{dzdaM} we are only considering variations in the parameters $z$ and $\alpha$ while keeping $\nu$ and $\ell_3$ fixed. Indeed the first law in the form \eqref{1stlaw2} does not hold for other variations. However, one may consider, in the spirit of `extended black hole thermodynamics' \cite{Kubiznak:2012wp}, identifying new `work' terms so that a more general first law holds. This would then be an extended thermodynamics for the quantum BTZ black hole. Presumably it should reduce in the limit of zero backreaction to the classical form in \cite{Frassino:2015oca}.

\paragraph{Charge.} The AdS C-metric has an extension to include electric or magnetic charge of the bulk black hole \cite{Plebanski:1976gy}. The metric induced on a Karch-Randall brane is not that of the charged BTZ black hole in Einstein-Maxwell theory \cite{Martinez:1999qi}, but it is nevertheless another valid solution, with a bulk structure similar to that of the Reissner-Nordstrom black hole. Much of the analysis in this paper can be extended to these solutions, which should yield another class of holographic quantum black holes.

\paragraph{Entanglement islands.}  Finally, let us mention that since the backreaction of the quantum fields is captured here in an explicit, analytic solution, the holographic construction of quBTZ black holes can be used to study detailed aspects of entanglement islands, following the ideas in \cite{Almheiri:2019hni}.

\bigskip

In the future we hope to report on at least some of these problems.

\section*{Acknowledgments}

We thank Veronika Hubeny, Don Marolf, Mukund Rangamani, and Marija Tomašević for useful discussions and suggestions. We are especially grateful to Mukund Rangamani for generously cooperating in verifying the correctness of some of our results. Work supported by ERC Advanced Grant GravBHs-692951, MEC grant FPA2016-76005-C2-2-P, and AGAUR grant 2017-SGR 754.

\newpage

\appendix

\section{Glossary of main symbols}\label{app:list}

\textit{Physical quantities}
\medskip

\noindent
\begin{tabular}{lp{0.7\textwidth}l}
$\ell_{4}$:& AdS$_{4}$ radius in the bulk &\eqref{l34}\\
$\ell_{3}$:& AdS$_{3}$ radius on the brane ($L_3$ + higher curvature corrections) &\eqref{emptyads4}\\
$\ell$:& brane position &\eqref{changeco}\\
& inverse of brane tension &\eqref{tension}\\
&strength of backreaction &\eqref{smallnu3d}\\ 
&cutoff length of 3D effective theory &\eqref{effact} \eqref{lggLpl}\\
$L_{3}$:& cosmological constant in 3D effective theory &\eqref{L3ell3} \\
$G_3$:& Newton's constant in 3D effective theory &\eqref{G34}\\ 
$\mathcal{G}_{3}$:& `renormalized' 3D Newton constant (accounting for higher curvature corrections to the mass) &\eqref{approxG} \eqref{exactG}\\
$c$:& central charge of CFT$_3$ &\eqref{central} \eqref{nu3d}\\
$S_\text{gen}$:& generalized entropy of quantum black hole (horizon area in bulk)&\eqref{quS} \eqref{bulkS}\\
$S_\text{out}$:&entanglement entropy of quantum fields outside black hole&\eqref{quS}\\
$S_\text{cl}$:&Bekenstein-Hawking entropy (horizon area on brane)&\eqref{Sclst}\\
$S_W$:&Bekenstein-Hawking-Wald entropy&\eqref{SW}\\
$S_\text{BTZ}$:& entropy of BTZ black hole with mass $M$ and spin $J$&\eqref{limBTZ}\\
\end{tabular}
\\

\bigskip
\noindent\textit{Auxiliary parameters}
\medskip

\noindent
\begin{tabular}{lp{0.68\textwidth}l}
$\kappa$:&small ($\kappa=+1$) or large ($\kappa=-1$) AdS$_3$ quantum black hole &\eqref{b1a} \eqref{b1b}\\
$\mu$:& quantum corrections parameter &\eqref{FM}\\
$x_1$:& smallest positive root of $G(x)$, axis of $\partial_\phi$ &\eqref{xrange} \eqref{mux1} \eqref{mux$1a$}\\
&black hole mass parameter &\eqref{Mstat} \eqref{MquBTZ}\\
$\Delta$:&periodicity of $\phi$&\eqref{Delx1} \eqref{Deltax$1a$}\\
$r_+$: &largest positive root of $H(r)$, horizon position&\\
$z$:& black hole mass parameter &\eqref{zdef}\\
$\nu$:&backreaction parameter&\eqref{nudef}\\
$a$, $\tilde{a}$, $\alpha$:& rotation and spin parameters&\eqref{rotc} \eqref{atilde} \eqref{alphadef}\\
\end{tabular}

\section{Limit of no backreaction}\label{app:zerobr}

Here we show that taking the limit $\ell\to 0$ in the AdS C-metric we recover a double Wick rotation of the Kerr-AdS$_4$ solution. This was the construction employed in \cite{Hubeny:2009rc} to obtain a four-dimensional bulk solution with a BTZ black hole at its asymptotic boundary. As explained in the main text, in this limit the brane tension becomes infinite and the brane moves towards the asymptotic AdS$_4$ boundary, where there appears a non-dynamical, classical BTZ geometry. The holographic stress tensor is that of the dual CFT in this fixed background.

Sending $\ell\to 0$ in \eqref{rotc} we find
\beqa
\label{rotczerol}
ds^2=\frac{\ell^2}{x^2 r^2}&\Biggl[& -\frac{H(r)}{\Sigma(x,r)} \lp dt +a x^2d\phi \rp^2+
\frac{\Sigma(x,r)}{H(r)}dr^2\nn\\
&&+r^2\lp \frac{\Sigma(x,r)}{G(x)}dx^2+\frac{G(x)}{\Sigma(x,r)}\lp d\phi-\frac{a}{r^2}dt\rp^2\rp
\Biggr]\,,
\eeqa
where
\beq
H(r)= \frac{r^2}{\ell_3^2}+\kappa +\frac{a^2}{r^2}\,,
\eeq
while $G(x)$ and $\Sigma(x,r)$ remain unchanged. The bulk cosmological constant $\ell_4=\ell$ appears as an overall prefactor which we rescale so as to keep it finite. We can easily recognize that when $\kappa=-1$ the metric induced at the boundary at $x\to 0$, where $G=\Sigma=1$, is conformally equivalent to a rotating BTZ black hole (the bulk global structure is explained in sec.~\ref{subsec:geoqbtz}).

Now let us perform a double Wick rotation of \eqref{rotczerol} by transforming coordinates
\begin{align}
t&=i\ell_3\frac{\sqrt{1+\hat{a}^2}}{1-\hat{a}^2}\hat{\Phi}\,, & \phi & =i \frac{\sqrt{1+\hat{a}^2}}{1-\hat{a}^2}\lp \hat{a}\hat{\Phi} -\hat{T}\rp\,,\nn\\
r & =\ell_3\frac{\hat{a}}{\sqrt{1+\hat{a}^2}}\frac1{\hat{X}}\,, &
x&=\sqrt{1+\hat{a}^2}\frac1{\hat{R}}\,,
\end{align}
and parameters
\beq
a=\ell_3\frac{\hat{a}}{1+\hat{a}^2}\,,\qquad \mu =\frac{\hat{\mu}}{\lp 1+\hat{a}^2\rp^{3/2}}\,.
\eeq
We also set $\kappa=-1$.
The metric \eqref{rotczerol} then takes the form
\begin{equation}\label{KerrAdS2}
\mathrm ds^2=\ell_4^2\left[-\frac{\Delta_R}{\xi^2\Xi^2}\left(\mathrm d\hat T-\frac{\hat a^2-\hat X^2}{\hat a}\mathrm d\hat\Phi\right)^2+\frac{\Delta_X}{\xi^2\Xi^2}\left(\mathrm d\hat T-\frac{\hat a^2+\hat R^2}{\hat a}\mathrm d\hat\Phi\right)^2+
\xi^2\lp \frac{d\hat X^2}{\Delta_X}+\frac{d\hat R^2}{\Delta_R}\rp\right]\,,
\end{equation}
where
\begin{align}
\Delta_R&=(\hat a^2+\hat R^2)(1+\hat R^2)-\hat\mu\hat R\,, &\Delta_X&=(\hat a^2-\hat X^2)(1-\hat X^2)\,,\nn\\
\xi&=\sqrt{\hat X^2+\hat R^2}\,,& \Xi&=1-\hat a^2\,.
\end{align}
This is the Kerr-AdS$_4$ solution in a coordinate system where $\hat R$ and $\hat X$ are treated on nearly equal footing.  It is still a two-parameter family given by $\hat \mu$ and $\hat a$, and has an overall scale $\ell_4$. We can bring it to a more conventional form by further changing
\begin{align}
\hat{T}&=\lp 1-\frac{\mathsf{a}^2}{\ell_4^2}\rp\frac{T}{\ell_4}\,, & \hat{\Phi}&=\Phi\,,\nn\\
\hat{R} &= \frac{\rho}{\ell_4}\,, &\hat{X}&=\frac{\mathsf{a}}{\ell_4}\cos\Theta\,,
\end{align}
with
\beq
\hat{\mu}=\frac{2m}{\ell_4}\,,\qquad \hat{a}=\frac{\mathsf{a}}{\ell_4}\,.
\eeq
Then \eqref{KerrAdS2} becomes
\beq\label{KerrAdS}
ds^2 = -\frac{\Delta _{\rho }}{\zeta^2}\left(dT-\frac{\mathsf{a}}{\Xi } \sin ^2\Theta \, d\Phi \right)^2+
\frac{\Delta _{\Theta } \sin ^2\Theta}{\zeta^2}\left(\mathsf{a} dT-\frac{\mathsf{a}^2+\rho ^2}{\Xi }d\Phi \right)^2+
\frac{ \zeta^2}{\Delta _{\Theta }}d\Theta^2+
\frac{\zeta^2}{\Delta _{\rho }}d\rho^2\,,
\eeq
with metric functions
\begin{align}
\label{RotAdS4}
\Delta _{\rho}(\rho)&=\left(\mathsf{a}^2+\rho ^2\right) \left(\frac{\rho ^2}{\ell_4^2}+1\right)-2m  \rho\,, &
\Delta _{\Theta }(\Theta)&=1-\frac{\mathsf{a}^2}{\ell_4^2} \cos^2 \Theta\,,\nn\\
 \zeta (\rho,\Theta)&=\sqrt{\rho ^2+\mathsf{a}^2 \cos^2 \Theta}\,, & \Xi& =1-\frac{\mathsf{a}^2}{\ell_4^2}\,,
\end{align}
and
\beq
m= \frac{1}{2\rho_{+}}(\rho_{+}^{2}+\mathsf{a}^2)\left(1+\frac{\rho_{+}^{2}}{\ell_4^2} \right)\,.
\eeq
This form of the Kerr-AdS$_4$ metric was the starting point in \cite{Hubeny:2009rc}, which we have then proven is a limit of the construction in this paper.

\section{Stress tensor in BTZ parameters}\label{app:TabBTZ}

The holographic stress tensor can be usefully rewritten in what we refer to as `BTZ parameters' $\mathsf{r}_{\pm}$, defined in terms of $M$ and $J$ as
\beq
\mathsf{r}_{\pm}=\frac{\ell_3}{2}\lp \sqrt{8\mathcal{G}_3\lp M+\frac{J}{\ell_3}\rp} \pm \sqrt{8\mathcal{G}_3\lp M-\frac{J}{\ell_3}\rp}\rp \,,
\eeq
or equivalently
\bea
8 \mathcal{G}_3 M &=& \frac{\mathsf{r}_{+}^2+\mathsf{r}_{-}^2}{\ell_3^2}\,,\\
8 \mathcal{G}_3 J &=& \frac{2 \mathsf{r}_{+} \mathsf{r}_{-}}{\ell_3}\,.
\eea
These correspond to the radii of the outer and inner horizons in the BTZ black hole, which is the zero backreaction limit $\ell\to 0$ of quBTZ \eqref{rotquBTZ}. When $\ell\neq 0$ the quantum backreaction to the metric displaces these horizons away from their classical position (in terms of $\mathcal{G}_3 M$ and $\mathcal{G}_3 J$), and in that case $\mathsf{r}_{\pm}$ are merely a convenient reparametrization of $\mathcal{G}_3 M$ and $\mathcal{G}_3 J$ and not horizon radii.

In principle it is possible to express the dependence of the stress tensor on $M$ and $J$ in terms of only $(\mathsf{r}_+,\mathsf{r}_-)$, but since it involves these in roots of cubics, there are limits to how much the final expressions can be simplified. We deal with this by introducing (yet another) two new auxiliary parameters, $\xi$ and $\tilde{\alpha}$, most simply defined as
\beq
-\kappa  x_{1}^2 = \left(1+\tilde{\alpha}^2\right) \xi,
\qquad
\frac{a x_{1}^2}{\ell_{3}} = \tilde{\alpha}\, \xi
\eeq
and which, using \eqref{MquBTZ} and \eqref{JquBTZ}, are related to $\mathsf{r}_{\pm}$ by
\bea
\mathsf{r}_{+} &=& \ell_3\frac{2 \sqrt{\xi } \,  \left(1+\tilde{\alpha}^2 \xi\right)}{3 +\xi+\tilde{\alpha}^2 (1-\xi ) \xi }\,,\\
\mathsf{r}_{-} &=&  \ell_3\frac{2 \tilde{\alpha} \sqrt{\xi }\, (1+\xi)}{3 +\xi+\tilde{\alpha}^2 (1-\xi ) \xi }\,.
\eea
In \eqref{Ttt}-\eqref{Trr} the stress tensor is written in terms of $r$, which is related to the BTZ radial coordinate $\bar{r}$ as in \eqref{rtorbar}.
Making all the changes, the stress tensor takes the form
\begin{align}
8\pi G_3\langle T^{\bar{t}}{}_{\bar{t}} \rangle&=
g(\mathsf{r}_{\pm})
\left[ (\mathsf{r}_{+}^2+2\mathsf{r}_{-}^2)(\bar{r}^2-\mathsf{r}_{-}^2)-3 \tilde{\alpha}\, \mathsf{r}_{-} \mathsf{r}_{+} \left(2 \bar{r}^2-\mathsf{r}_{-}^2-\mathsf{r}_{+}^2\right) +\tilde{\alpha}^2 \left(\bar{r}^2-\mathsf{r}_{+}^2\right) \left(\mathsf{r}_{-}^2+2 \mathsf{r}_{+}^2\right)
\right],\label{Ttt2}\\
8\pi G_3\langle T^{\bar{\phi}}{}_{\bar{\phi}} \rangle&=
-g(\mathsf{r}_{\pm})
\left[ (2 \mathsf{r}_{+}^2 + \mathsf{r}_{-}^2)(\bar{r}^2-\mathsf{r}_{-}^2)  - 3 \tilde{\alpha}\, \mathsf{r}_{-} \mathsf{r}_{+} \left(2 \bar{r}^2-\mathsf{r}_{-}^2-\mathsf{r}_{+}^2\right)+
\tilde{\alpha}^2 \left(\bar{r}^2-\mathsf{r}_{+}^2\right) \left(2 \mathsf{r}_{-}^2+\mathsf{r}_{+}^2\right)
\right]\,,\\
8\pi G_3\langle T^{\bar{t}}{}_{\bar{\phi}} \rangle&=
-3 \ell_3 g(\mathsf{r}_{\pm}) \mathsf{r}_{-} \mathsf{r}_{+}
\left[ \bar{r}^2-\mathsf{r}_{+}^2+\frac{\tilde{\alpha} \left(2 \mathsf{r}_{-}^2 \mathsf{r}_{+}^2-\bar{r}^2 \left(\mathsf{r}_{-}^2+\mathsf{r}_{+}^2\right)\right)}{\mathsf{r}_{-} \mathsf{r}_{+} }+\tilde{\alpha}^2 \left(\bar{r}^2-\mathsf{r}_{+}^2\right)
\right]\,,\\
8\pi G_3\langle T^{\bar{\phi}}{}_{\bar{t}} \rangle&=
\frac{3  g(\mathsf{r}_{\pm})\mathsf{r}_{-} \mathsf{r}_{+}}{\ell_3}
\left[
\bar{r}^2-\mathsf{r}_{-}^2+\frac{\tilde{\alpha} \left(\mathsf{r}_{-}^4+\mathsf{r}_{+}^4-\bar{r}^2 \left(\mathsf{r}_{-}^2+\mathsf{r}_{+}^2\right)\right)}{\mathsf{r}_{-} \mathsf{r}_{+}}+\tilde{\alpha}^2 \left(\bar{r}^2-\mathsf{r}_{+}^2\right)
\right]\,,\\
8\pi G_3\langle T^{\bar{r}}{}_{\bar{r}} \rangle&=
g(\mathsf{r}_{\pm})
\left(\mathsf{r}_{+}^2-\mathsf{r}_{-}^2\right) \left(\bar{r}^2-\mathsf{r}_{-}^2-\tilde{\alpha}^2 \left(\bar{r}^2-\mathsf{r}_{+}^2\right)\right)
\label{Trr2}\,,
\end{align}
where
\beq
g(\mathsf{r}_{\pm}) = \frac{\ell}{2 \ell_3^3}
\frac{
1+\left(1+\tilde{\alpha}^2\right) \xi +\tilde{\alpha}^2 \xi ^2}{ \xi ^{3/2} }
\frac{ \sqrt{\mathsf{r}_{+}^2-\mathsf{r}_{-}^2}}{\left(\bar{r}^2-\mathsf{r}_{-}^2-\tilde{\alpha}^2(\bar{r}^2-\mathsf{r}_{+}^2)\right)^{5/2}}\,.
\eeq
The complicated, implicit dependence on $(\mathsf{r}_+,\mathsf{r}_-)$ of this last function is a reflection of how the bulk geometry holographically solves for the properties of the CFT. The rest of the stress tensor depends on $(\mathsf{r}_+,\mathsf{r}_-)$ in a cleaner way, and in this form it is possible to compare this dependence to that of the coefficients in \eqref{fTtt}-\eqref{fTrr} for free fields \cite{Casals:2019jfo}. We leave this exercise to the interested reader.

Observe that the value of $\bar{r}$ at the ring singularity \eqref{ringbarr} that the metric \eqref{rotquBTZ} has when $\ell\neq 0$, is given by
\beq
r_s^2=\frac{\mathsf{r}_{-}^2-\tilde{\alpha}^2\mathsf{r}_{+}^2}{1-\tilde{\alpha}^2}\,,
\eeq
and this is the only place where the stress tensor becomes singular.


\begin{thebibliography}{99}
\bibliographystyle{JHEP-2}
\bibliography{QSTinBTZ}



\bibitem{Fabbri:2005mw}
A.~Fabbri and J.~Navarro-Salas,
``Modeling black hole evaporation,''
Imp.~Coll.~Press, London, UK (2005).

\bibitem{Callan:1992rs}
C.~G.~Callan, Jr., S.~B.~Giddings, J.~A.~Harvey and A.~Strominger,
``Evanescent black holes,''
Phys. Rev. D \textbf{45} (1992) no.4, 1005
doi:10.1103/PhysRevD.45.R1005
[arXiv:hep-th/9111056 [hep-th]].

\bibitem{Almheiri:2019psf}
A.~Almheiri, N.~Engelhardt, D.~Marolf and H.~Maxfield,
``The entropy of bulk quantum fields and the entanglement wedge of an evaporating black hole,''
JHEP \textbf{12} (2019), 063
doi:10.1007/JHEP12(2019)063
[arXiv:1905.08762 [hep-th]].

\bibitem{Banados:1992wn}
  M.~Ba{\~n}ados, C.~Teitelboim and J.~Zanelli,
  ``The Black hole in three-dimensional space-time,''
  Phys.\ Rev.\ Lett.\  {\bf 69} (1992) 1849
  doi:10.1103/PhysRevLett.69.1849
  [hep-th/9204099].

\bibitem{Banados:1992gq}
  M.~Ba{\~n}ados, M.~Henneaux, C.~Teitelboim and J.~Zanelli,
  ``Geometry of the (2+1) black hole,''
  Phys.\ Rev.\ D {\bf 48} (1993) 1506
   Erratum: [Phys.\ Rev.\ D {\bf 88} (2013) 069902]
  doi:10.1103/PhysRevD.48.1506, 10.1103/PhysRevD.88.069902
  [gr-qc/9302012].



\bibitem{Randall:1999vf}
L.~Randall and R.~Sundrum,
``An Alternative to compactification,''
Phys. Rev. Lett. \textbf{83} (1999), 4690-4693
doi:10.1103/PhysRevLett.83.4690
[arXiv:hep-th/9906064 [hep-th]].

\bibitem{Karch:2000ct}
  A.~Karch and L.~Randall,
  ``Locally localized gravity,''
  JHEP {\bf 0105} (2001) 008
  doi:10.1088/1126-6708/2001/05/008
  [hep-th/0011156].

\bibitem{Verlinde:1999fy}
  H.~L.~Verlinde,
  ``Holography and compactification,''
  Nucl.\ Phys.\ B {\bf 580} (2000) 264
  doi:10.1016/S0550-3213(00)00224-8
  [hep-th/9906182].

\bibitem{Gubser:1999vj}
  S.~S.~Gubser,
  ``AdS/CFT and gravity,''
  Phys.\ Rev.\ D {\bf 63} (2001) 084017
  doi:10.1103/PhysRevD.63.084017
  [hep-th/9912001].

\bibitem{Geng:2020qvw}
H.~Geng and A.~Karch,
``Massive islands,''
JHEP \textbf{09} (2020), 121
doi:10.1007/JHEP09(2020)121
[arXiv:2006.02438 [hep-th]].

\bibitem{Chen:2020uac}
H.~Z.~Chen, R.~C.~Myers, D.~Neuenfeld, I.~A.~Reyes and J.~Sandor,
``Quantum Extremal Islands Made Easy, Part I: Entanglement on the Brane,''
[arXiv:2006.04851 [hep-th]].


\bibitem{Emparan:2002px}
  R.~Emparan, A.~Fabbri and N.~Kaloper,
  ``Quantum black holes as holograms in AdS brane worlds,''
  JHEP {\bf 0208} (2002) 043
  doi:10.1088/1126-6708/2002/08/043
  [hep-th/0206155].

\bibitem{Emparan:1999fd}
  R.~Emparan, G.~T.~Horowitz and R.~C.~Myers,
  ``Exact description of black holes on branes. 2. Comparison with BTZ black holes and black strings,''
  JHEP {\bf 0001} (2000) 021
  doi:10.1088/1126-6708/2000/01/021
  [hep-th/9912135].

\bibitem{Tanaka:2002rb}
T.~Tanaka,
``Classical black hole evaporation in Randall-Sundrum infinite brane world,''
Prog. Theor. Phys. Suppl. \textbf{148} (2003), 307-316
doi:10.1143/PTPS.148.307
[arXiv:gr-qc/0203082 [gr-qc]].


\bibitem{Steif:1993zv}
  A.~R.~Steif,
  ``The Quantum stress tensor in the three-dimensional black hole,''
  Phys.\ Rev.\ D {\bf 49} (1994) 585
  doi:10.1103/PhysRevD.49.R585
  [gr-qc/9308032].

\bibitem{Shiraishi:1993nu}
K.~Shiraishi and T.~Maki,
``Vacuum polarization around a three-dimensional black hole,''
Class. Quant. Grav. \textbf{11} (1994), 695-700
doi:10.1088/0264-9381/11/3/019
[arXiv:1505.03958 [gr-qc]].

\bibitem{Shiraishi:2018pdw}
K.~Shiraishi and T.~Maki,
``Quantum fluctuation of stress tensor and black holes in three dimensions,''
Phys. Rev. D \textbf{49} (1994), 5286-5294
doi:10.1103/PhysRevD.49.5286
[arXiv:1804.07872 [gr-qc]].

\bibitem{Lifschytz:1993eb}
G.~Lifschytz and M.~Ortiz,
``Scalar field quantization on the (2+1)-dimensional black hole background,''
Phys. Rev. D \textbf{49} (1994), 1929-1943
doi:10.1103/PhysRevD.49.1929
[arXiv:gr-qc/9310008 [gr-qc]].

\bibitem{Martinez:1996uv}
C.~Martinez and J.~Zanelli,
``Back reaction of a conformal field on a three-dimensional black hole,''
Phys. Rev. D \textbf{55} (1997), 3642-3646
doi:10.1103/PhysRevD.55.3642
[arXiv:gr-qc/9610050 [gr-qc]].

\bibitem{Casals:2016odj}
  M.~Casals, A.~Fabbri, C.~Martínez and J.~Zanelli,
  ``Quantum Backreaction on Three-Dimensional Black Holes and Naked Singularities,''
  Phys.\ Rev.\ Lett.\  {\bf 118} (2017) no.13,  131102
  doi:10.1103/PhysRevLett.118.131102
  [arXiv:1608.05366 [gr-qc]].

\bibitem{Casals:2019jfo}
  M.~Casals, A.~Fabbri, C.~Martínez and J.~Zanelli,
  ``Quantum-corrected rotating black holes and naked singularities in (2+1) dimensions,''
  Phys.\ Rev.\ D {\bf 99} (2019) no.10,  104023
  doi:10.1103/PhysRevD.99.104023
  [arXiv:1902.01583 [hep-th]].


\bibitem{Hubeny:2009rc}
  V.~E.~Hubeny, D.~Marolf and M.~Rangamani,
  ``Hawking radiation from AdS black holes,''
  Class.\ Quant.\ Grav.\  {\bf 27} (2010) 095018
  doi:10.1088/0264-9381/27/9/095018
  [arXiv:0911.4144 [hep-th]].

\bibitem{Wald:1993nt}
R.~M.~Wald,
``Black hole entropy is the Noether charge,''
Phys. Rev. D \textbf{48} (1993) no.8, 3427-3431
doi:10.1103/PhysRevD.48.R3427
[arXiv:gr-qc/9307038 [gr-qc]].

\bibitem{Ryu:2006bv}
S.~Ryu and T.~Takayanagi,
``Holographic derivation of entanglement entropy from AdS/CFT,''
Phys. Rev. Lett. \textbf{96} (2006), 181602
doi:10.1103/PhysRevLett.96.181602
[arXiv:hep-th/0603001 [hep-th]].

\bibitem{Emparan:2006ni}
R.~Emparan,
``Black hole entropy as entanglement entropy: A Holographic derivation,''
JHEP \textbf{06} (2006), 012
doi:10.1088/1126-6708/2006/06/012
[arXiv:hep-th/0603081 [hep-th]].

\bibitem{Engelhardt:2014gca}
N.~Engelhardt and A.~C.~Wall,
``Quantum Extremal Surfaces: Holographic Entanglement Entropy beyond the Classical Regime,''
JHEP \textbf{01} (2015), 073
doi:10.1007/JHEP01(2015)073
[arXiv:1408.3203 [hep-th]].

\bibitem{Almheiri:2019hni}
A.~Almheiri, R.~Mahajan, J.~Maldacena and Y.~Zhao,
``The Page curve of Hawking radiation from semiclassical geometry,''
JHEP \textbf{03} (2020), 149
doi:10.1007/JHEP03(2020)149
[arXiv:1908.10996 [hep-th]].

\bibitem{Almheiri:2019psy}
A.~Almheiri, R.~Mahajan and J.~E.~Santos,
``Entanglement islands in higher dimensions,''
SciPost Phys. \textbf{9} (2020) no.1, 001
doi:10.21468/SciPostPhys.9.1.001
[arXiv:1911.09666 [hep-th]].



\bibitem{Emparan:1999wa}
R.~Emparan, G.~T.~Horowitz and R.~C.~Myers,
``Exact description of black holes on branes,''
JHEP \textbf{01} (2000), 007
doi:10.1088/1126-6708/2000/01/007
[arXiv:hep-th/9911043 [hep-th]].

\bibitem{Plebanski:1976gy}
J.~Plebanski and M.~Demianski,
``Rotating, charged, and uniformly accelerating mass in general relativity,''
Annals Phys. \textbf{98} (1976), 98-127
doi:10.1016/0003-4916(76)90240-2

\bibitem{deHaro:2000wj}
S.~de Haro, K.~Skenderis and S.~N.~Solodukhin,
``Gravity in warped compactifications and the holographic stress tensor,''
Class. Quant. Grav. \textbf{18} (2001), 3171-3180
doi:10.1088/0264-9381/18/16/307
[arXiv:hep-th/0011230 [hep-th]].

\bibitem{Emparan:1999pm}
R.~Emparan, C.~V.~Johnson and R.~C.~Myers,
``Surface terms as counterterms in the AdS / CFT correspondence,''
Phys. Rev. D \textbf{60} (1999), 104001
doi:10.1103/PhysRevD.60.104001
[arXiv:hep-th/9903238 [hep-th]].

\bibitem{Bergshoeff:2009hq}
E.~A.~Bergshoeff, O.~Hohm and P.~K.~Townsend,
``Massive Gravity in Three Dimensions,''
Phys. Rev. Lett. \textbf{102} (2009), 201301
doi:10.1103/PhysRevLett.102.201301
[arXiv:0901.1766 [hep-th]].


\bibitem{Dvali:2000hr}
G.~R.~Dvali, G.~Gabadadze and M.~Porrati,
``4-D gravity on a brane in 5-D Minkowski space,''
Phys. Lett. B \textbf{485} (2000), 208-214
doi:10.1016/S0370-2693(00)00669-9
[arXiv:hep-th/0005016 [hep-th]].

\bibitem{Aharony:2008ug}
O.~Aharony, O.~Bergman, D.~L.~Jafferis and J.~Maldacena,
``N=6 superconformal Chern-Simons-matter theories, M2-branes and their gravity duals,''
JHEP \textbf{10} (2008), 091
doi:10.1088/1126-6708/2008/10/091
[arXiv:0806.1218 [hep-th]].

\bibitem{Dvali:2007hz}
G.~Dvali,
``Black Holes and Large N Species Solution to the Hierarchy Problem,''
Fortsch. Phys. \textbf{58} (2010), 528-536
doi:10.1002/prop.201000009
[arXiv:0706.2050 [hep-th]].


\bibitem{Deser:1981wh}
S.~Deser, R.~Jackiw and S.~Templeton,
``Topologically Massive Gauge Theories,''
Annals Phys. \textbf{140} (1982), 372-411
doi:10.1016/0003-4916(82)90164-6

\bibitem{Bergshoeff:2009aq}
E.~A.~Bergshoeff, O.~Hohm and P.~K.~Townsend,
``More on Massive 3D Gravity,''
Phys. Rev. D \textbf{79} (2009), 124042
doi:10.1103/PhysRevD.79.124042
[arXiv:0905.1259 [hep-th]].


\bibitem{Jacobson:1993vj}
T.~Jacobson, G.~Kang and R.~C.~Myers,
``On black hole entropy,''
Phys. Rev. D \textbf{49} (1994), 6587-6598
doi:10.1103/PhysRevD.49.6587
[arXiv:gr-qc/9312023 [gr-qc]].

\bibitem{Cremonini:2009ih}
S.~Cremonini, J.~T.~Liu and P.~Szepietowski,
``Higher Derivative Corrections to R-charged Black Holes: Boundary Counterterms and the Mass-Charge Relation,''
JHEP \textbf{03} (2010), 042
doi:10.1007/JHEP03(2010)042
[arXiv:0910.5159 [hep-th]].

\bibitem{Marolf:2013ioa}
D.~Marolf, M.~Rangamani and T.~Wiseman,
``Holographic thermal field theory on curved spacetimes,''
Class. Quant. Grav. \textbf{31} (2014), 063001
doi:10.1088/0264-9381/31/6/063001
[arXiv:1312.0612 [hep-th]].

\bibitem{Dias:2019ery}
  O.~J.~C.~Dias, H.~S.~Reall and J.~E.~Santos,
  ``The BTZ black hole violates strong cosmic censorship,''
  JHEP {\bf 1912} (2019) 097
  doi:10.1007/JHEP12(2019)097
  [arXiv:1906.08265 [hep-th]].

\bibitem{Hollands:2019whz}
S.~Hollands, R.~M.~Wald and J.~Zahn,
``Quantum Instability of the Cauchy Horizon in Reissner-Nordstr\"om-deSitter Spacetime,''
Class. Quant. Grav. \textbf{37} (2020) no.11, 115009
doi:10.1088/1361-6382/ab8052
[arXiv:1912.06047 [gr-qc]].


\bibitem{Emparan:2020rnp}
R.~Emparan and M.~Tomašević,
``Strong cosmic censorship in the BTZ black hole,''
JHEP \textbf{06} (2020), 038
doi:10.1007/JHEP06(2020)038
[arXiv:2002.02083 [hep-th]].



\bibitem{Papadodimas:2019msp}
  K.~Papadodimas, S.~Raju and P.~Shrivastava,
  ``A simple quantum test for smooth horizons,''
  arXiv:1910.02992 [hep-th].

\bibitem{Balasubramanian:2019qwk}
V.~Balasubramanian, A.~Kar and G.~Sárosi,
``Holographic Probes of Inner Horizons,''
JHEP \textbf{06} (2020), 054
doi:10.1007/JHEP06(2020)054
[arXiv:1911.12413 [hep-th]].



\bibitem{Emparan:2000rs}
R.~Emparan, G.~T.~Horowitz and R.~C.~Myers,
``Black holes radiate mainly on the brane,''
Phys. Rev. Lett. \textbf{85} (2000), 499-502
doi:10.1103/PhysRevLett.85.499
[arXiv:hep-th/0003118 [hep-th]].

\bibitem{Anabalon:2018ydc}
A.~Anabalón, M.~Appels, R.~Gregory, D.~Kubizňák, R.~B.~Mann and A.~Ovgün,
``Holographic Thermodynamics of Accelerating Black Holes,''
Phys. Rev. D \textbf{98} (2018) no.10, 104038
doi:10.1103/PhysRevD.98.104038
[arXiv:1805.02687 [hep-th]].



\bibitem{Wall:2011hj}
A.~C.~Wall,
``A proof of the generalized second law for rapidly changing fields and arbitrary horizon slices,''
Phys. Rev. D \textbf{85} (2012), 104049
doi:10.1103/PhysRevD.85.104049
[arXiv:1105.3445 [gr-qc]].

\bibitem{Figueras:2011gd}
P.~Figueras and T.~Wiseman,
``Gravity and large black holes in Randall-Sundrum II braneworlds,''
Phys. Rev. Lett. \textbf{107} (2011), 081101
doi:10.1103/PhysRevLett.107.081101
[arXiv:1105.2558 [hep-th]].


\bibitem{Kubiznak:2012wp}
D.~Kubiznak and R.~B.~Mann,
``P-V criticality of charged AdS black holes,''
JHEP \textbf{07} (2012), 033
doi:10.1007/JHEP07(2012)033
[arXiv:1205.0559 [hep-th]].

\bibitem{Frassino:2015oca}
A.~M.~Frassino, R.~B.~Mann and J.~R.~Mureika,
``Lower-Dimensional Black Hole Chemistry,''
Phys. Rev. D \textbf{92} (2015) no.12, 124069
doi:10.1103/PhysRevD.92.124069
[arXiv:1509.05481 [gr-qc]].

\bibitem{Martinez:1999qi}
C.~Martinez, C.~Teitelboim and J.~Zanelli,
``Charged rotating black hole in three space-time dimensions,''
Phys. Rev. D \textbf{61} (2000), 104013
doi:10.1103/PhysRevD.61.104013
[arXiv:hep-th/9912259 [hep-th]].

\end{thebibliography}
\end{document}